\documentclass[journal ]{new-aiaa}
\usepackage{amsmath}
\usepackage[utf8]{inputenc}
\usepackage{textcomp}
\usepackage{graphicx}
\usepackage[version=4]{mhchem}
\usepackage{siunitx}
\usepackage{longtable,tabularx}
\usepackage{hyperref}
\usepackage{latexsym}
\usepackage{multirow}
\usepackage{makecell}
\usepackage{graphicx}
\usepackage{float}
\usepackage{subfigure}
\usepackage{subcaption}

\usepackage{algorithm}
\usepackage{algpseudocode}
\title{Identifying Fixed Points in the Three-Body Problem Using a High-Order Transfer Map}

\author{Xingyu Zhou\footnote{Ph.D. Candidate, School of Aerospace Engineering; \url{zhouxingyu@bit.edu.cn}.}}
\affil{Beijing Institute of Technology, 100081 Beijing, People’s Republic of China}

\author{Lorenzo Anoè\footnote{Ph.D. Candidate, Te Pūnaha Ātea - Space Institute, 20 Symonds Street, Auckland Central; \url{lorenzo.anoe@gmail.com}.} and Roberto Armellin\footnote{Professor, Te Pūnaha Ātea - Space Institute, 20 Symonds Street, Auckland Central; \url{roberto.armellin@auckland.ac.nz} (Member AIAA).}}
\affil{University of Auckland, Auckland 1010, New Zealand}

\author{Dong Qiao\footnote{Professor, School of Aerospace Engineering; \url{qiaodong@bit.edu.cn} (Corresponding Author).} and Xiangyu Li\footnote{Associate Professor, School of Aerospace Engineering; \url{lixiangy@bit.edu.cn}.}}
\affil{Beijing Institute of Technology, 100081 Beijing, People’s Republic of China}

\begin{document}

\maketitle

\begin{abstract}
Periodic orbits (POs) play a central role in the circular restricted three-body problem (CRTBP). 
This paper introduces a method to search for POs by identifying single- and multiple-revolution fixed points in chosen Poincaré maps that describe the CRTBP dynamics, with a theoretical capability to detect all fixed points across arbitrary revolution counts exhaustively.
First, high-order transfer maps (HOTMs), represented as polynomials, are constructed within the differential algebra (DA) framework for both planar and spatial CRTBP to map states between successive Poincaré section crossings, with the Jacobi constant used to reduce the number of independent variables. 
Next, an automatic domain splitting (ADS) strategy is employed to generate subdomains, preserving HOTM accuracy, with an integrated feasibility estimation to reduce ADS’s computation burden.
Then, a two-stage HOTM-based polynomial optimization framework is introduced, first identifying combinable subdomain sequences and then refining the fixed point solutions. 
Finally, the method is applied to the Earth-Moon CRTBP, identifying POs up to nine revolutions in the planar case and four in the spatial case. Known families such as distant retrograde orbits (DROs) and Lyapunov orbits are recovered, along with a previously undocumented family that exhibits a hybrid character between DROs and Lyapunov orbits. 
\end{abstract}

%%%%%%%%%%%%%%%%%%%%%%%%%%%%%%%%%%%%%%%%%%%%%%%%%%%%%%%%%%%%%%%%%%%%%%%%%%%%%%%%%%%%%%%%%%%
%%%%%%%%%%%%%%%%%%%%%%%%%%%%%%%%%%%%%%%%%%%%%%%%%%%%%%%%%%%%%%%%%%%%%%%%%%%%%%%%%%%%%%%%%%%
\section{Introduction}
\lettrine{T}{hree-body} space exploration, particularly in the Sun-Earth and Earth-Moon systems, has garnered increasing interest in recent years \cite{muralidharan2023stretching}. These regions offer wide and stable viewpoints for scientific observation \cite{Gardner2006, Zhou2022SST}, allowing for low-energy transfers due to their unique dynamical structures \cite{Anderson2009, Anderson2012, Parker2013, qi2020achievable}. The NASA ARTEMIS program plans to use a near-rectilinear halo orbit (NRHO) for the Lunar Gateway space station \cite{Smith2020, batcha2020artemis}, and the CNSA Chang'e program has demonstrated growing capabilities in Earth-Moon exploration \cite{Gao2019AAST, wang2021orbit}. Three-body space is now a key area for deploying observatories, supporting crewed lunar missions, and enabling future deep space exploration \cite{Qi2025}.

Within these mission frameworks, periodic orbits (POs) in the three-body problem have emerged as fundamental structures for designing trajectories and conducting long-term operations \cite{Grebow2006, Fu2022ASR, Li2025SCPMA}. In the context of dynamical systems, a PO is a closed trajectory that returns to its initial state after a finite period; this initial state is a fixed point of the dynamical flow \cite{broucke1968periodic, Koh2017}. 
For example, the NRHO targeted by the Lunar Gateway is part of the halo L2 family of POs, which bifurcates from the Lyapunov L2 family \cite{Smith2020}.
More broadly, POs organize the phase space and provide natural pathways for transfers \cite{Anderson2014, Anderson2020, Pushparaj2021, Sullivan2023, Caleb2024}, station-keeping \cite{Fu2022ASR}, and capture \cite{Li2019JGCD, Anoe2024JGCD, Anoe2025BC3DandLTB, anoè2025optimization, Chaudhary2024, liangutilization}. Their associated invariant manifolds enable low-energy connections between distant regions, support ballistic capture strategies, and define the geometry of stability regions. As such, POs serve not only as operational platforms but also as gateways for accessing broader dynamical structures in three-body dynamical environments.

Most studies of POs dynamics are conducted within the framework of the Circular Restricted Three-Body Problem (CRTBP) \cite{Grebow2006, Baoyin2006, McKay2011, Yang2023SST}, a fundamental model that captures the essential gravitational dynamics between two massive primaries and a massless third body \cite{Szebehely1967}. The CRTBP provides a rich dynamical environment while remaining analytically and computationally tractable, making it widely used for theoretical analysis and mission design. Under this model, a variety of PO families, such as Lyapunov \cite{petropoulos2004low}, halo \cite{FARQUHAR1976, breakwell1979halo, connor1984three}, and distant retrograde orbits (DROs) \cite{bezrouk2017long, yin2023midcourse, Cui2025JGCD, Wang2025Astro}, have been extensively investigated. These orbits, together with their stable and unstable invariant manifolds, have been used to construct low-energy trajectories connecting libration point regions with the surfaces of the primaries. For instance, previous work has demonstrated how manifolds of Lyapunov orbits near the Sun-Earth and Earth-Moon L1/L2 points can naturally guide ballistic transfers to planetary surfaces or Earth parking orbits, enabling sample return and staging missions with minimal fuel cost \cite{Baoyin2006}.

Within the CRTBP, numerous techniques have been developed to compute POs, trace their continuation along families, and analyze bifurcations. Among these, collocation-based methods \cite{olikara2016computation} have recently gained attention for their robustness in computing complex dynamical structures such as quasi-periodic tori and heteroclinic connections. The numerical continuation package AUTO constitutes an alternative approach that approximates POs via piecewise polynomial representations, and it has a longer history in this context \cite{dichmann2003computation, doedel2003computation}. Unlike shooting methods, which rely on trajectory integration and do not yield an explicit functional form of the orbit, AUTO constructs a segmented polynomial approximation of the solution, enabling accurate computation and continuation of POs. This method has been successfully applied to the CRTBP to compute extensive families of POs and to study their bifurcation structures in detail \cite{dichmann2003computation, doedel2003computation}.
Foundational work by Llibre and Simó \cite{llibre1980oscillatory} also played a crucial role in the numerical exploration of the CRTBP, including the first numerical approximation of complex solutions such as Chenciner's figure-8 orbit.

To effectively utilize POs in space exploration missions, it is first necessary to identify them within the chosen dynamical model. In most cases, this process reduces to locating fixed points on a Poincaré section \cite{Winter2000, capdevila2014transfer}. However, in the CRTBP model, closed-form solutions for such fixed points do not exist. As a result, numerical methods play a central role in the search for POs under the CRTBP model. Classical techniques, such as Newton’s method, have been widely used for this purpose \cite{Grebow2006, Marchand2007}. However, these methods rely heavily on high-quality initial guesses to ensure convergence. Without a sufficiently accurate starting point, the iteration may diverge or converge to an irrelevant solution \cite{qi2018study}. Even when a good initial guess is available, the method typically yields only one fixed point, often corresponding to a single-revolution orbit, while failing to uncover the complete set of periodic solutions, including higher-period or multi-revolution POs. These limitations highlight the need for more robust and global methods for fixed-point computation in the CRTBP.

An alternative to direct root-finding is the fixed-point search method of Wittig and Berz \cite{wittig2014high}, which starts from a prescribed domain and uses high-order polynomial enclosures to approximate the mapped set. By testing for intersections with the original domain and recursively bisecting it, regions guaranteed not to contain fixed points are discarded, while others are refined until the desired precision is reached. This method has been applied to identify fixed points in simplified models of the Tevatron accelerator \cite{wittig2014high}.
Although this method offers rigorous guarantees and global coverage, it is not without limitations. It does not directly compute fixed points and often requires extensive subdivision to ensure accuracy. Moreover, for multiple-revolution fixed points, it requires propagating the polynomial map across the whole period, rather than reusing and composing a single-period polynomial map, which leads to additional computational overhead. Moreover, the accelerator model considered is not Hamiltonian, unlike the CRTBP studied in this work, but rather closer to a dissipative system.

Motivated by the work of Wittig and Berz \cite{wittig2014high}, this paper proposes a polynomial-based method for identifying both single- and multiple-revolution fixed points in the CRTBP. While both approaches aim to identify high-period fixed points exhaustively, the proposed method constructs only a single-revolution polynomial map, which is reused to obtain multiple-revolution fixed points. Initially, building on the high-order transfer map (HOTM) techniques originally developed for perturbed Keplerian motion \cite{Wittig2015CMDA}, this work extends their application to both planar and spatial CRTBP scenarios. The HOTMs are constructed within the differential algebra (DA) framework. Leveraging the properties of the CRTBP, the Jacobi constant and Poincaré section are imposed as constraints in the HOTMs to reduce the dimensionality of the fixed-point search. An automatic domain splitting (ADS) tool is applied to control the approximation error of HOTMs. Unlike Wittig and Berz’s method \cite{wittig2014high}, the splitting here is not used to locate fixed points directly but to ensure the accuracy of HOTMs. 
Feasibility identification is incorporated during the splitting process, eliminating infeasible subdomains to reduce computational load.
Then, a two-stage polynomial optimization framework is introduced to identify fixed points. The first stage identifies combinable subdomain sequences, enabling the repeated use of a single-revolution HOTM for multi-revolution analysis. The second stage refines these sequences to locate fixed points directly through polynomial optimization, without relying on domain subdivision. The performance of the proposed method in terms of searching fixed points will be validated through two Earth-Moon CRTBP cases, which identify the POs up to 9 revolutions in the planar case and 4 in the spatial case.

The remainder of this paper is organized as follows. Section~\ref{sec:Preliminaries} briefly describes the CRTBP, DA, and ADS. Section~\ref{sec:High-Order Transfer Map} first derives the HOTMs for both planar and spatial CRTBP cases, followed by an introduction to the feasibility estimation-based ADS process. The two-stage polynomial optimization framework for identifying fixed points is presented in Sec.~\ref{sec:Identifying Fixed Points Using Polynomial Optimizatio}. Numerical examples are given in Sec.~\ref{sec:Numerical Examples}, and conclusions are provided in Sec.~\ref{sec:Conclusion}.

%%%%%%%%%%%%%%%%%%%%%%%%%%%%%%%%%%%%%%%%%%%%%%%%%%%%%%%%%%%%%%%%%%%%%%%%%%%%%%%%%%%%%%%%%%%
%%%%%%%%%%%%%%%%%%%%%%%%%%%%%%%%%%%%%%%%%%%%%%%%%%%%%%%%%%%%%%%%%%%%%%%%%%%%%%%%%%%%%%%%%%%
\section{Preliminaries} \label{sec:Preliminaries}

%%%%%%%%%%%%%%%%%%%%%%%%%%%%%%%%%%%%%%%%%%%%%%%%%%%%%%%%%%%%%%%%%%%%%%%%%%%%%%%%%%%%%%%%%%%
\subsection{Circular Restricted Three-Body Problem} \label{sec:Circular Restricted Three-Body Problem}
The CRTBP is a simplified model of the general three-body problem, in which two massive primary bodies move in circular orbits around their common barycenter, and a third body of negligible mass moves under their gravitational influence. The CRTBP model is formulated in a rotating coordinate frame where the primaries remain fixed along the \emph{x}-axis: the more massive primary is located at $x=-\mu$, and the less massive one at $x=1-\mu$, where $\mu  = \frac{{{m_2}}}{{{m_1} + {m_2}}}$ is the mass ratio between two primaries ($m_1$ and $m_2$ correspond to the mass of the more massive and the less massive primaries, respectively). The CRTBP equations of motion are written as
\begin{equation} \label{eq1}
    \left\{ \begin{array}{l}
        \ddot x = 2\dot y + x - \frac{{(1 - \mu )(x + \mu )}}{{r_1^3}} - \frac{{\mu (x + \mu  - 1)}}{{r_2^3}}\\
        \ddot y =  - 2\dot x + y - \frac{{(1 - \mu )y}}{{r_1^3}} - \frac{{\mu y}}{{r_2^3}}\\
        \ddot z =  - \frac{{(1 - \mu )z}}{{r_1^3}} - \frac{{\mu z}}{{r_2^3}}
    \end{array} \right. \,,
\end{equation}
where $\boldsymbol{x} = [\boldsymbol{r};\boldsymbol{v}] = [x,y,z,\dot x,\dot y,\dot z] \in {\mathbb{R}^6}$ is the nondimensional orbital state in the rotating frame, and
\begin{equation} \label{eq2}
    {r_1} = \sqrt {{{(x + \mu )}^2} + {y^2} + {z^2}} \,,
\end{equation}
\begin{equation} \label{eq3}
    {r_2} = \sqrt {{{(x + \mu  - 1)}^2} + {y^2} + {z^2}} \,.
\end{equation}

The CRTBP model described above corresponds to the spatial (three-dimensional) case, where the third body is allowed to move in all three directions $(x,y,z)$. The spatial CRTBP is symmetric with respect to the \emph{x}-\emph{y} plane (\emph{i.e.}, the $z=0$ plane). A common simplification of the model is the planar CRTBP, in which the motion is restricted to the \emph{x}-\emph{y} plane and the state vector in Eq.~\eqref{eq1} becomes $\boldsymbol{x} = [\boldsymbol{r};\boldsymbol{v}] = [x,y,\dot x,\dot y] \in {\mathbb{R}^4}$. This occurs when the initial position and velocity components in the \emph{z}-direction are zero: $z = \dot z = 0$, and they remain zero at all times due to this symmetry. The planar CRTBP is widely used for analysis and mission design when out-of-plane motion is negligible.

An important integral of motion in the CRTBP is the Jacobi constant, $C_J$, which is a conserved total three-body energy in the rotating reference frame. 
It is defined as
\begin{equation} \label{eq4}
    C_J = 2U(x,y,z) - ({\dot x^2} + {\dot y^2} + {\dot z^2}) \,,
\end{equation}
where $U(x,y,z) = \frac{1}{2}({x^2} + {y^2}) + \frac{{1 - \mu }}{{{r_1}}} + \frac{\mu }{{{r_2}}}$ is the effective potential.

In this work, the CRTBP model is adopted due to its balance between analytical tractability and dynamical richness. 
Moreover, families of POs can only be described in the CRTBP, as its Hamiltonian nature allows for the description of families parametrised by the Jacobi constant. Well-known PO families, such as Lyapunov orbits, halo orbits, and DROs, exist only within the CRTBP framework. 
In more realistic, high-fidelity ephemeris models, these orbits do not remain strictly periodic, and corrections to initial conditions can yield quasi-periodic or bounded trajectories.

%%%%%%%%%%%%%%%%%%%%%%%%%%%%%%%%%%%%%%%%%%%%%%%%%%%%%%%%%%%%%%%%%%%%%%%%%%%%%%%%%%%%%%%%%%%
\subsection{Differential Algebra} \label{sec:Differential Algebra}
The DA technique was originally developed as an algebraic approach to tackle analytical problems \cite{DA_book}. It enables the derivation of high-order Taylor expansions for the flow of nonlinear orbital dynamics with respect to system parameters, such as the initial state \cite{Armellin2009CMDA, Armellin2012CMDA, Fossa2024} or maneuvers \cite{Zhou2025RS}. Unlike traditional numerical methods that evaluate functions at discrete points, DA manipulates functions as algebraic objects by expanding them into Taylor polynomials, thereby preserving local behavior and supporting operations such as composition, inversion, differentiation, and integration \cite{Valli2013}.

For a given nonlinear mapping $\boldsymbol{z} = \boldsymbol{F}(\boldsymbol{y})$, where $\boldsymbol{y} = {[{y_1},{y_2}, \cdots ,{y_m}]^{\rm{T}}} \in {\mathbb{R}^m}$ represents a vector of independent input variables, the DA variable is constructed by introducing a deviation $\delta \boldsymbol{y}$ around a nominal reference point $\boldsymbol{\bar y}$; that is, $[\boldsymbol{y}] = \boldsymbol{\bar y} + \delta \boldsymbol{y}$. This expression represents a Taylor expansion around the reference point $\boldsymbol{\bar y}$. Substituting the DA variable $[\boldsymbol{y}]$ into the nonlinear function $\boldsymbol{z} = \boldsymbol{F}(\boldsymbol{y})$, an \emph{n}-th order Taylor polynomial expansion (\emph{n} is a user-defined parameter) of the output $\boldsymbol{z}$ with respect to the perturbation $\delta \boldsymbol{y}$ can be obtained as
\begin{equation} \label{eq8}
    \begin{aligned}
        {[\boldsymbol{z}]} & =\boldsymbol{F}([\boldsymbol{y}])=\boldsymbol{F}(\bar{\boldsymbol{y}}+\delta \boldsymbol{y}) \\ & =\mathscr{T}_{\boldsymbol{z}}(\delta \boldsymbol{y})=\sum_{k_1+\cdots+k_m \leq n} \boldsymbol{c}_{k_1, \cdots, k_m} \cdot \delta y_1^{k_1} \cdots \delta y_m^{k_m}
    \end{aligned} \,,
\end{equation}
where ${\boldsymbol{c}_{{k_1}, \cdots ,{k_m}}}$ are the polynomial coefficients. In Eq.~\eqref{eq8} and throughout this paper, the symbol $\mathscr{T}$ ($T$ in script font) denotes a general high-order Taylor polynomial map.
The input variables $\delta \boldsymbol{y}$ in the map described by Eq.~\eqref{eq8} are placed inside the parentheses, while the map's output is indicated by the subscript $\boldsymbol{z}$.

The Differential Algebra Core Engine (DACE) \footnote{Available at \url{https://github.com/dacelib/dace}.} is a widely used and efficient C++ toolbox for DA computation. In this work, the DA computations are implemented using DACEyPy \footnote{Available at \url{https://github.com/giovannipurpura/daceypy}.}, the Python interface of the DACE library.

%%%%%%%%%%%%%%%%%%%%%%%%%%%%%%%%%%%%%%%%%%%%%%%%%%%%%%%%%%%%%%%%%%%%%%%%%%%%%%%%%%%%%%%%%%%
\subsection{Automatic Domain Splitting} \label{sec:Automatic Domain Split}
The DA approach efficiently replaces numerous pointwise integrations, as in Monte Carlo simulations, with high-order Taylor polynomial representations of orbital dynamics \cite{Valli2013, Fu2024Astro}. Yet in strongly nonlinear regimes like the three-body problem, a single expansion may lose accuracy, with convergence deteriorating along sensitive directions due to local instabilities. To address this issue, the ADS technique was introduced \cite{Wittig2015}. It monitors the decay of polynomial coefficients and adaptively subdivides the domain when accuracy falls below a threshold, generating new Taylor expansions in each subregion. This yields localized models that maintain reliability under strong nonlinearities and support accurate trajectory modeling in complex regimes such as the CRTBP \cite{Wittig2015, Evans2024}.

%%%%%%%%%%%%%%%%%%%%%%%%%%%%%%%%%%%%%%%%%%%%%%%%%%%%%%%%%%%%%%%%%%%%%%%%%%%%%%%%%%%%%%%%%%%
\subsection{Differential Algebra-Based Time Expansion} \label{sec:Differential Algebra-Based Time Expansion}
To implement the time expansion in the DA framework, an artificial propagation variable $\tau  \in [0,1]$ is introduced \cite{Sun2019AA, Fu2024Astro, Zhou2025RS}. The physical time $t$ is then expressed as a linear function of $\tau$, given by $t = {\rm{ToF}} \cdot \tau  + {t_0}$,
where $\rm{ToF}$ is the total time-of-flight and $t_0$ is the initial time. The state $\boldsymbol{y}$ is then propagated with respect to the artificial propagation variable $\tau$, treating the $\rm{ToF}$ as a DA variable. This results in the following augmented system:
\begin{equation} \label{eq11}
    \left\{ {\begin{array}{*{20}{l}}
    {\cfrac{{d\boldsymbol{y}}}{{d\tau }} = {\rm{ToF}} \cdot \cfrac{d \boldsymbol{y}}{dt}}\\
    {\cfrac{{d{\rm{ToF}}}}{{d\tau }} = 0}
    \end{array}} \right. \,,
\end{equation}
where $\frac{d \boldsymbol{y}}{dt}$ denotes the derivative of the state $\boldsymbol{y}$ with respect to the physical time $t$, with the CRTBP model in Eq.~\eqref{eq1} being a special case. The second equation simply states that $\rm{ToF}$ remains constant during the expansion. By integrating the augmented system in Eq.~\eqref{eq11} over the interval $\tau \in [0,1]$ in the DA framework, the state $\boldsymbol{y}$ becomes a polynomial: ${\boldsymbol{y}} \approx {\mathscr {T}}_{\boldsymbol{y}}(\delta {\boldsymbol{y}_0},\delta {\rm{ToF}})$, where $\delta {\boldsymbol{y}_0} \in {\mathbb{R}^m}$ represents the deviation of the initial state $\boldsymbol{y}_0=\boldsymbol{y}(t_0) \in {\mathbb{R}^m}$, and $\delta\rm{ToF}$ is the variation of the $\rm{ToF}$. 

%%%%%%%%%%%%%%%%%%%%%%%%%%%%%%%%%%%%%%%%%%%%%%%%%%%%%%%%%%%%%%%%%%%%%%%%%%%%%%%%%%%%%%%%%%%
%%%%%%%%%%%%%%%%%%%%%%%%%%%%%%%%%%%%%%%%%%%%%%%%%%%%%%%%%%%%%%%%%%%%%%%%%%%%%%%%%%%%%%%%%%%
\section{High-Order Transfer Map} \label{sec:High-Order Transfer Map}
To improve the efficiency of constructing HOTMs, it is desirable to reduce the number of variables involved in the high-order Taylor polynomial expansion. In this study, we achieve the reduction by introducing two constraints: a Poincaré section constraint ($y=0$, $\dot{y} \geq 0$, and $0<x<1-\mu$) and a Jacobi constant constraint. These constraints effectively eliminate two degrees of freedom from the system, one associated with the final state lying on a predefined section, and another enforcing energy consistency, thus simplifying the polynomial representation of the HOTMs. In the following subsections, we present the methodology for generating HOTMs from a Poincaré section back to itself, separately for the planar and spatial formulations of the CRTBP.

%%%%%%%%%%%%%%%%%%%%%%%%%%%%%%%%%%%%%%%%%%%%%%%%%%%%%%%%%%%%%%%%%%%%%%%%%%%%%%%%%%%%%%%%%%%
\subsection{Planar case} \label{sec:Planar CRTBP case}
\begin{figure}[!h]
    \centering
    \includegraphics[width=0.5\linewidth]{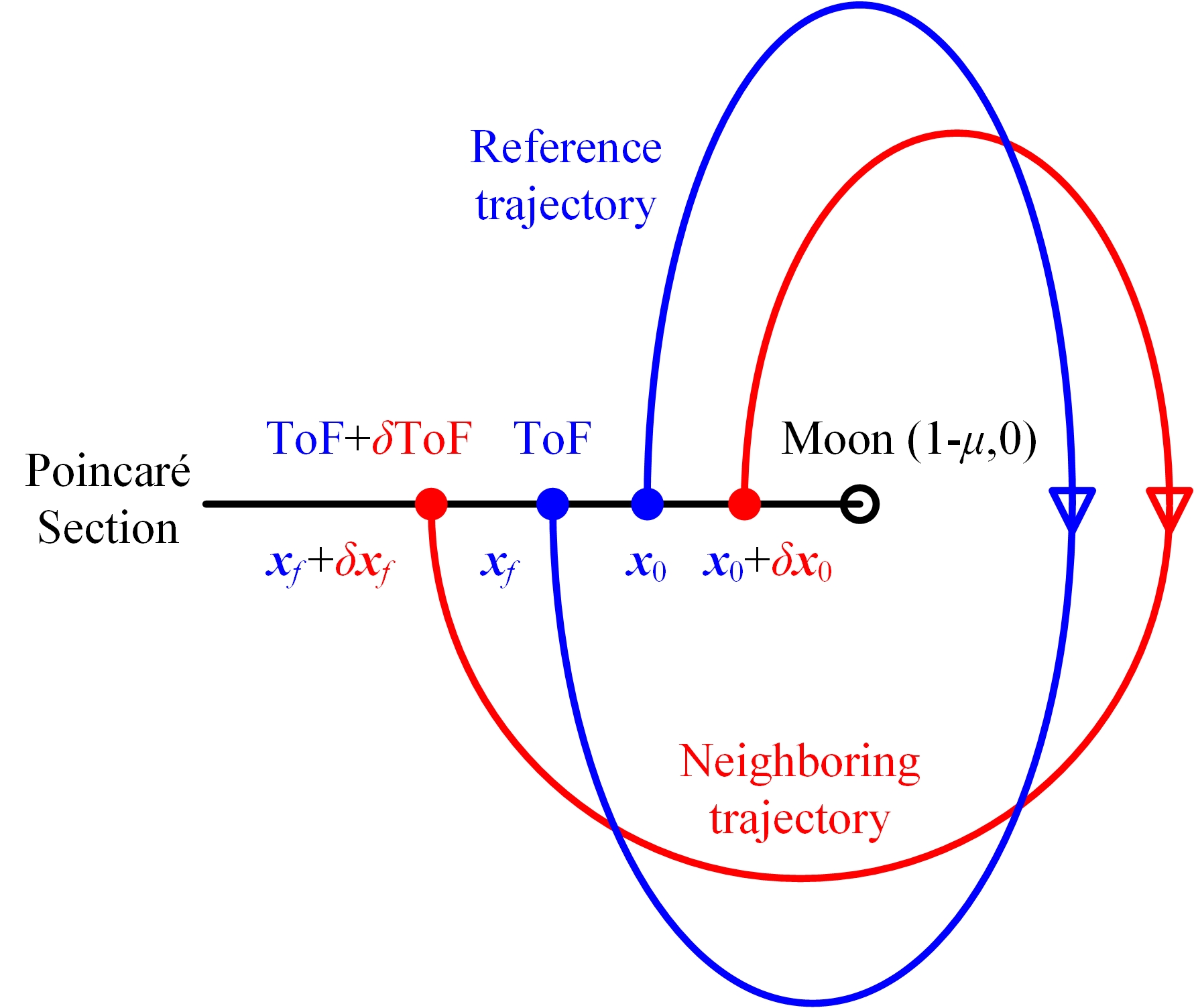}
    \caption{\label{fig1} An illustration of the reference and neighboring trajectories.}
\end{figure}

Figure~\ref{fig1} illustrates the concept of HOTM construction using a reference trajectory. The blue curve represents the reference trajectory, while the red curve shows a neighboring trajectory initialized with a small perturbation in the initial condition. The black horizontal line indicates the Poincaré section, which both trajectories intersect. The objective of the HOTM developed in this section is to approximate the return state of the neighboring trajectory on the Poincaré section, specifically, its state at the second intersection after departing from the same section.

Let ${\boldsymbol{x}_0} = {[{x_0},{y_0} = 0,{\dot x_0},{\dot y_0}\geq 0]^T} \in {\mathbb{R}^4}$ be a reference state on the selected Poincaré section, and $C_J$ be the corresponding Jacobi constant. By perturbing the reference state in the DA framework, one has
\begin{equation} \label{eq13}
    [{\boldsymbol{x}_0}] = {\boldsymbol{x}_0} + \delta {\boldsymbol{x}_0} = \left( {\begin{array}{*{20}{c}}
        {{x_0} + \delta {x_0}}\\
        0\\
        {{{\dot x}_0} + \delta {{\dot x}_0}}\\
        {{{\dot y}_0} + \delta {{\dot y}_0}}
    \end{array}} \right) \,,
\end{equation}
where $[{\boldsymbol{x}_0}]$ represents a perturbed initial state defined in the DA framework, and each component of the perturbed initial state $[{\boldsymbol{x}_0}]$ is written as $[x_0]={x_0} + \delta {x_0}$, $[{\dot x}_0]={\dot x}_0 + \delta {{\dot x}_0}$, and $[{\dot y}_0]={\dot y}_0 + \delta {{\dot y}_0}$.
Then, with a given fixed value of $C_J$, $\delta {\dot y_0}$ can be analytically derived within the DA framework, given as
\begin{equation} \label{eq14}
    \delta {\dot y_0} \approx {{\mathscr T}_{\delta {{\dot y}_0}}}(\delta {x_0},\delta {\dot x_0}) 
    = - \dot{y}_0+\sqrt{2U(x_0+\delta x_0,y=0,z=0)-(\dot{x}_0+\delta \dot{x}_0)^2-C_J} \,,
\end{equation}
which shows that the ${\dot y_0}$ degree of freedom can be removed under the Jacobi constant constraint. Then, the perturbation term $\delta {\boldsymbol{x}_0}$ in Eq.~\eqref{eq13} can be rewritten as
\begin{equation} \label{eq15}
    \delta {\boldsymbol{x}_0} = \left( {\begin{array}{*{20}{c}}
        {{{\mathscr I}_{\delta {x_0}}}}\\
        0\\
        {{{\mathscr I}_{\delta {{\dot x}_0}}}}\\
        {{{\mathscr T}_{\delta {{\dot y}_0}}}}
        \end{array}} \right) \left( {\begin{array}{*{20}{c}}
        {\delta {x_0}}\\
        {\delta {{\dot x}_0}}
    \end{array}} \right) \,,
\end{equation}
where ${\mathscr I}_{\delta {x_0}}(\delta {x_0})=\delta {x_0}$ and ${\mathscr I}_{\delta {{\dot x}_0}}(\delta {{\dot x}_0})=\delta {{\dot x}_0}$ are identity maps.

Let $\mathrm{ToF}$ be the nominal time-of-flight (corresponding to the reference state $\boldsymbol{x}_0$) from the chosen Poincaré section back to itself, which serves as the reference propagation interval for constructing the HOTM. Perturbing the time-of-flight as $\mathrm{ToF}+\delta \mathrm{ToF}$ and applying the time expansion described in Sec.~\ref{sec:Differential Algebra-Based Time Expansion}, one has
\begin{equation} \label{eq16}
    \delta {\boldsymbol{x}_f} = \left( {\begin{array}{*{20}{c}}
        {\delta {x_f}}\\
        {\delta {y_f}}\\
        {\delta {{\dot x}_f}}\\
        {\delta {{\dot y}_f}}
        \end{array}} \right) \approx \left( {\begin{array}{*{20}{c}}
        {{{\mathscr T}_{\delta {x_f}}}}\\
        {{{\mathscr T}_{\delta {y_f}}}}\\
        {{{\mathscr T}_{\delta {{\dot x}_f}}}}\\
        {{{\mathscr T}_{\delta {{\dot y}_f}}}}
        \end{array}} \right) \left( {\begin{array}{*{20}{c}}
        {\delta {x_0}}\\
        {\delta {{\dot x}_0}}\\
        {\delta {\rm{ToF}}}
    \end{array}} \right) \,,
\end{equation}
which is a polynomial approximation of the final state deviation $\delta {\boldsymbol{x}_f}$ (the subscript `\emph{f}' represents final) with respect to $\delta {x_0}$, $\delta {\dot x_0}$, and $\delta {\rm{ToF}}$. 
By introducing two identified maps $\delta {x_0} = {{\mathscr I}_{\delta {x_0}}}(\delta {x_0})$ and $\delta {\dot x_0} = {{\mathscr I}_{\delta {{\dot x}_0}}}(\delta {\dot x_0})$, the map $\delta {y_f} = {{\mathscr T}_{\delta {y_f}}}(\delta {x_0}, \delta {\dot{x}_0}, \delta \mathrm{ToF})$ can be augmented as
\begin{equation} \label{eq17}
    \left( {\begin{array}{*{20}{c}}
        {\delta {y_f}}\\
        {\delta {x_0}}\\
        {\delta {{\dot x}_0}}
        \end{array}} \right) \approx \left( {\begin{array}{*{20}{c}}
        {{{\mathscr T}_{\delta {y_f}}}}\\
        {{{\mathscr I}_{\delta {x_0}}}}\\
        {{{\mathscr I}_{\delta {{\dot x}_0}}}}
        \end{array}} \right) \left( {\begin{array}{*{20}{c}}
        {\delta {x_0}}\\
        {\delta {{\dot x}_0}}\\
        {\delta {\rm{ToF}}}
    \end{array}} \right) \,.
\end{equation}

Then, inverting the map in Eq.~\eqref{eq17} to obtain
\begin{equation} \label{eq18}
    \left( {\begin{array}{*{20}{c}}
        {\delta {x_0}}\\
        {\delta {{\dot x}_0}}\\
        {\delta {\rm{ToF}}}
    \end{array}} \right) \approx \left( {\begin{array}{*{20}{c}}
        {{{\mathscr T}_{\delta {y_f}}}}\\
        {{{\mathscr I}_{\delta {x_0}}}}\\
        {{{\mathscr I}_{\delta {{\dot x}_0}}}}
        \end{array}} \right)^{-1} \left( {\begin{array}{*{20}{c}}
        {\delta {y_f}}\\
        {\delta {x_0}}\\
        {\delta {{\dot x}_0}}
        \end{array}} \right) \, ,
\end{equation}
and applying $\delta {y_f} = 0$ (the Poincaré section constraint) to the map \eqref{eq18}, yields to
\begin{equation} \label{eq19}
    \delta {\rm{ToF}} \approx {{\mathscr T}_{\delta {\rm{ToF}}}}(\delta {x_0},\delta {\dot x_0},\delta {y_f} = 0) = {{\mathscr T}_{\delta {\rm{ToF}}}}(\delta {x_0},\delta {\dot x_0}) \,.
\end{equation}
Finally, substituting the polynomial in Eq.~\eqref{eq19} back into Eq.~\eqref{eq16}, one has
\begin{equation} \label{eq20}
    \delta {\boldsymbol{x}_f} = \left( {\begin{array}{*{20}{c}}
        {\delta {x_f}}\\
        {\delta {y_f}}\\
        {\delta {{\dot x}_f}}\\
        {\delta {{\dot y}_f}}
        \end{array}} \right) \approx \left( {\begin{array}{*{20}{c}}
        {{{\mathscr T}_{\delta {x_f}}}}\\
        0\\
        {{{\mathscr T}_{\delta {{\dot x}_f}}}}\\
        {{{\mathscr T}_{\delta {{\dot y}_f}}}}
        \end{array}} \right) \left( {\begin{array}{*{20}{c}}
        {\delta {x_0}}\\
        {\delta {{\dot x}_0}}
    \end{array}} \right) \,,
\end{equation}
which is the HOTM of the planar case. 
Since the Poincaré section condition imposes $\delta y_f = 0$, and the Jacobi constant further constrains $\delta \dot y_f$ as a function of $\delta x_f$ and $\delta \dot x_f$, only two components of the final state are independent.
Accordingly, within the HOTM, we only need to consider the output components $\delta x_f$ and $\delta \dot x_f$:
\begin{equation} \label{eq21}
    \left( {\begin{array}{*{20}{c}}
        {\delta {x_f}}\\
        {\delta {{\dot x}_f}}
        \end{array}} \right) \approx \left( {\begin{array}{*{20}{c}}
        {{{\mathscr T}_{\delta {x_f}}}}\\
        {{{\mathscr T}_{\delta {{\dot x}_f}}}}
        \end{array}} \right) \left( {\begin{array}{*{20}{c}}
        {\delta {x_0}}\\
        {\delta {{\dot x}_0}}
    \end{array}} \right) \,.
\end{equation}

Algorithm~\ref{alg1} summarizes the procedure for constructing the HOTM in the planar CRTBP case. It is worth noting that there are two possible ways to determine the initial reference state: (1) one can prescribe the \emph{y}-axis velocity and compute the corresponding Jacobi constant $C_J$; or (2) one can prescribe the Jacobi constant $C_J$ and solve for the compatible ${\dot y_0}$ (see \textbf{step 1} in Algorithm~\ref{alg1}). In this work, the second approach is adopted. 
However, for a given Jacobi constant $C_J$, the value of ${\dot y_0}$ generally admits two real solutions, corresponding to the positive and negative square roots. In this work, we limit the analysis to ${\dot y_0} > 0$, and we construct the HOTM using only this condition.

\begin{algorithm}[!h]
    \caption{Pseudocode to construct the HOTM for the planar case.} \label{alg1}
    \begin{algorithmic}[1]
    \State{Determine the reference ${\dot y_0}$ using the Jacobi constant $C_J$ and the center point $({x_0},{\dot x_0})$.} \Comment{using Eq.~\eqref{eq4}}
    \State{Construct the reference state ${\boldsymbol{x}_0} = {[{x_0},0,{\dot x_0},{\dot y_0}]^T}$ and numerically propagate (no DA) it under the CRTBP model until the next crossing with the Poincaré section to obtain the nominal $\rm{ToF}$.}
    \State{Perturb the reference state ${\boldsymbol{x}_0}$ in the DA framework to obtain $[\boldsymbol{x}_0]=\boldsymbol{x}_0+\delta \boldsymbol{x}_0$.} \Comment{using Eq.~\eqref{eq13}}
    \State{Remove $\delta {\dot y_0}$ degree of freedom using state within the Poincaré section ($x_0 + \delta x_0$ and $\dot{x}_0+ \delta \dot{x}_0$) and the Jacobi constant constraint.} \Comment{using Eqs.~\eqref{eq14}-\eqref{eq15}}
    \State{Add the $\delta \rm{ToF}$ as a DA variable and propagate the perturbed state within the Poincaré section with time expansion (in DA) to obtain $\delta {\boldsymbol{x}_f}$ in a polynomial fashion.} \Comment{using Eq.~\eqref{eq16}}
    \State{Remove the $\delta \rm{ToF}$ degree of freedom via a partial map inversion (in DA) on $\delta {\boldsymbol{x}_f}$ with the Poincaré section constraint ($y_f=0$) and substitute the $\delta \rm{ToF}$ polynomial back into the $\delta {\boldsymbol{x}_f}$ polynomial.} \Comment{using Eqs.~\eqref{eq17}-\eqref{eq21}}
    \State{Return the polynomial expansion for $\delta {\boldsymbol{x}_f}$, which is the HOTM.} \Comment{using Eq.~\eqref{eq21}}
    \end{algorithmic}
\end{algorithm}

An initial (rectangular) domain, centered at the reference point $({x_0},{\dot x_0})$, is defined as:
\begin{equation} \label{eq22}
    {{\mathscr D}_0} = [{x_0} - \Delta {x_0},{x_0} + \Delta {x_0}] \times [{\dot x_0} - \Delta {\dot x_0},{\dot x_0} + \Delta {\dot x_0}]
\end{equation}
with $\Delta {x_0}$ and $\Delta {\dot x_0}$ representing the half-lengths along the ${x_0}$ and ${\dot x_0}$ directions, respectively. That is, $ - \Delta {x_0} \le \delta {x_0} \le \Delta {x_0}$ and $ - \Delta {\dot x_0} \le \delta {\dot x_0} \le \Delta {\dot x_0}$. These are user-defined parameters that determine the size of the domain around the reference point. In addition, since planar POs are typically symmetric with respect to the \emph{x}-axis, the velocity component ${\dot x_0}$ at the reference point is set to zero when defining the initial domain in the velocity direction.

The proposed method searches for fixed points within this domain. If the domain ${\mathscr D}_0$ is small, a single high-order polynomial is often sufficient for accurate approximation, but the downside is that few or no fixed points may lie within such a narrow region. On the other hand, a larger domain increases the likelihood of containing multiple solutions. Nevertheless, this typically exceeds the accuracy limit of a single polynomial, thereby requiring the use of the ADS technique, which will be introduced in Sec.~\ref{sec:Automatic Domain Split with Feasibility Detection}.

%%%%%%%%%%%%%%%%%%%%%%%%%%%%%%%%%%%%%%%%%%%%%%%%%%%%%%%%%%%%%%%%%%%%%%%%%%%%%%%%%%%%%%%%%%%
\subsection{Spatial case} \label{sec:Spatial CRTBP case}
In the spatial CRTBP case, the dynamics model is extended by including the position and velocity components in the \emph{z}-direction. As a result, the overall formulation operates in a six-dimensional state space. However, the procedure for constructing the HOTM remains fundamentally the same as in the planar case (see Algorithm~\ref{alg1}). The key steps, such as DA-based propagation, time expansion, and constraint enforcement, follow an analogous structure. Therefore, in this section, we omit the repeated derivations and focus only on the differences introduced by the additional dimensions.

In the spatial case, the perturbed initial state in Eq.~\eqref{eq13} is rewritten as
\begin{equation} \label{eq23}
    [{\boldsymbol{x}_0}] = {\boldsymbol{x}_0} + \delta {\boldsymbol{x}_0} = \left( {\begin{array}{*{20}{c}}
        {{x_0} + \delta {x_0}}\\
        0\\
        {{z_0} + \delta {z_0}}\\
        {{{\dot x}_0} + \delta {{\dot x}_0}}\\
        {{{\dot y}_0} + \delta {{\dot y}_0}}\\
        {{{\dot z}_0} + \delta {{\dot z}_0}}
    \end{array}} \right) \,.
\end{equation}

Using the Jacobi constant constraint and following the derivation in Eqs.~\eqref{eq14}-\eqref{eq15}, one has
\begin{equation} \label{eq24}
    \begin{aligned}
        \delta {\dot y_0} &\approx {{\mathscr T}_{\delta {{\dot y}_0}}}(\delta {x_0},\delta {\dot x_0},\delta {z_0},\delta {\dot z_0}) \\
        &= - \dot{y}_0+\sqrt{2U(x_0+\delta x_0,y=0,z_0+\delta z_0)-(\dot{x}_0+\delta \dot{x}_0)^2-(\dot{z}_0+\delta \dot{z}_0)^2-C_J}
    \end{aligned} \,.
\end{equation}
Moreover, using time expansion, polynomial inversion, and enforcing the Poincaré section condition, the $\rm{ToF}$ variable is approximated as $\delta {\rm{ToF}} \approx {{\mathscr T}_{\delta {\rm{ToF}}}}(\delta {x_0},\delta {{\dot x}_0},\delta {z_0},\delta {{\dot z}_0},\delta {y_f} = 0)$, which finally yields the HOTM for the spatial case:
\begin{equation} \label{eq26}
    \left( {\begin{array}{*{20}{c}}
    {\delta {x_f}}\\
    {\delta {{\dot x}_f}}\\
    {\delta {z_f}}\\
    {\delta {{\dot z}_f}}
    \end{array}} \right) \approx \left( {\begin{array}{*{20}{c}}
    {{{\mathscr T}_{\delta {x_f}}}}\\
    {{{\mathscr T}_{\delta {{\dot x}_f}}}}\\
    {{{\mathscr T}_{\delta {z_f}}}}\\
    {{{\mathscr T}_{\delta {{\dot z}_f}}}}
    \end{array}} \right) \left( {\begin{array}{*{20}{c}}
    {\delta {x_0}}\\
    {\delta {{\dot x}_0}}\\
    {\delta {z_0}}\\
    {\delta {{\dot z}_0}}
    \end{array}} \right) \,.
\end{equation}

In contrast to the planar case, the spatial HOTM involves four variables: $\delta {x_0}$, $\delta {\dot x_0}$, $\delta {z_0}$, and $\delta {\dot z_0}$, leading to a four-dimensional initial domain. This domain defines the region in which the fixed point search is performed. Due to the inherent symmetry of the CRTBP model with respect to the $z=0$ plane, any trajectory initialized at $({z_0},{\dot z_0})$ has a mirrored counterpart at $( - {z_0}, - {\dot z_0})$. As a result, it is not necessary to explicitly include both positive and negative values of ${z_0}$ and ${\dot z_0}$ in the domain. Without loss of generality, we restrict the initial region to one side of the symmetry plane, which simplifies the search and reduces redundant computation.

In this work, the initial domain of the spatial case is defined as
\begin{equation} \label{eq27}
    \begin{aligned}
        {{\mathscr D}_0} &= [{x_0} - \Delta {x_0},{x_0} + \Delta {x_0}] \times [{{\dot x}_0} - \Delta {{\dot x}_0},{{\dot x}_0} + \Delta {{\dot x}_0}]\\
        & \times [0,2\Delta {z_0}] \times [- \Delta {{\dot z}_0},+ \Delta {{\dot z}_0}]
    \end{aligned} \,.
\end{equation}
The initial domain is defined to include $z=0$ by setting $z_0 = \Delta z_0 > 0$, which, due to symmetry, ensures coverage of trajectories crossing the $z=0$ plane. Both signs of $\dot z$ are considered, \emph{i.e.}, $\dot z < 0$ and $\dot z > 0$.

%%%%%%%%%%%%%%%%%%%%%%%%%%%%%%%%%%%%%%%%%%%%%%%%%%%%%%%%%%%%%%%%%%%%%%%%%%%%%%%%%%%%%%%%%%%
\subsection{Automatic Domain Splitting with Feasibility Detection} \label{sec:Automatic Domain Split with Feasibility Detection}
The previous subsection presented the procedure for constructing a HOTM centered around a reference trajectory. However, due to the strong nonlinearities inherent in the three-body dynamics, the deviation between the reference and neighboring trajectories can grow rapidly, especially over the return time interval. As a result, the accuracy of a single high-order polynomial becomes difficult to guarantee over a large initial domain. To overcome this limitation, the ADS technique is employed in this work, which adaptively partitions the initial domain and constructs localized polynomial approximations to maintain accuracy throughout the region of interest.

A major drawback of the ADS is its computational cost, as it requires the construction of a large number of high-order polynomials, one for each subdomain. First, to reduce computational time, a parallel implementation is employed during each iteration of the ADS process, enabling independent subdomains to be processed concurrently. Second, to reduce the actual computational cost, a feasibility detection method is proposed to work in conjunction with the ADS. This method identifies subdomains that are unlikely to contain valid fixed points and excludes them from further splitting. By pruning such infeasible subdomains once detected, the number of required polynomials is significantly reduced.

In this work, four types of infeasible subdomains are identified and excluded:
\begin{enumerate}
    \item \label{infeasibility1} The reference trajectory (\emph{i.e.}, the center point of the subdomain) fails to return to the selected Poincaré section within the specified time interval, making it impossible to determine a valid reference $\mathrm{ToF}$ for map construction (reference $\mathrm{ToF}$ is illustrated in Fig.~\ref{fig1}). 
    \item \label{infeasibility2} The reference trajectory results in a collision with the Earth or Moon, or passes too close to either body. Such close approaches may cause the DA integration to lose accuracy or become numerically unstable.
    \item \label{infeasibility3} Under the Jacobi constant constraint in Eq.~\eqref{eq4}, no admissible trajectory exists within the region. This may occur, for example, when the velocity magnitude is too large, violating the energy condition imposed by the Jacobi constant.
    \item \label{infeasibility4} No trajectory within the subdomain returns to the initial domain. Consequently, the region cannot contain a fixed point by definition.
\end{enumerate}

The first two types of infeasibility are straightforward to detect. Given a maximum allowable time-of-flight, $\rm{ToF}_{\max}$, and minimum allowable distance from the Earth and Moon, $d_E^{\min }$ and $d_M^{\min }$, one can numerically propagate the reference trajectory (\emph{i.e.}, the center point of each subdomain) using a standard numerical integrator (without using the DA framework). If the trajectory fails to return to the selected Poincaré section within the specified $\rm{ToF}_{\max}$, or if its closest approach to either primary body falls below the predefined threshold: ${r_1}  \le d_E^{\min }$ and ${r_2}  \le d_M^{\min }$, the subdomain is flagged as infeasible and excluded from further consideration.

In addition, the third type of infeasibility can be identified by evaluating the Jacobi constant expression in Eq.~\eqref{eq4}. Specifically, by substituting the values of ${x_0}$, ${\dot x_0}$, ${z_0}$, and ${\dot z_0}$ (${x_0}$ and ${\dot x_0}$ for the planar case) into the Jacobi constant equation, one can isolate $\dot y_0^2$ term as
\begin{equation} \label{eq28}
    \dot y_0^2 = 2U({x_0},{y_0},{z_0}) - (\dot x_0^2 + \dot z_0^2) - {C_J}
\end{equation}
and determine its sign. If the resulting value from Eq.~\eqref{eq28} is positive (or equal to 0), a real-valued solution for ${\dot y_0}$ exists, indicating that at least one admissible trajectory in the subdomain satisfies the energy constraint. Conversely, if $\dot y_0^2 < 0$, the subdomain is physically infeasible under the given Jacobi constant.

For the final type of infeasibility, once the first three conditions have been ruled out, a HOTM is constructed for the considered subdomain following the procedure outlined in the previous section. After obtaining the polynomial representation of the HOTM, a Linear Dominated Bounder (LDB) method \cite{makino2005range, Armellin2009CMDA} \footnote{The LDB algorithm is already embedded in the DACEypy package.} is used to estimate the range of the return state on the Poincaré section. The LDB algorithm is designed to compute conservative enclosures of polynomial outputs over a given input domain. Specifically, it estimates the lower and upper bounds of each output variable independently. As a result, for the planar case, the result is a rectangle, and for the spatial case, it becomes a four-dimensional hyperrectangle. Although this method produces overestimated bounds and may introduce some conservatism (especially in cases with strong nonlinearity, variable coupling, and high-order polynomial terms), it is computationally efficient. Moreover, the overestimation property ensures that truly feasible subdomains are not mistakenly discarded during the pruning process. 
If the range estimated by the LDB does not intersect with any feasible subdomain, the current subdomain is classified as infeasible and excluded from further processing. Here, the term feasible subdomain refers to those identified in the previous ADS iteration. Since intersections with infeasible subdomains cannot contain fixed points either, this strategy further reduces the number of subdomains to be considered, thereby lowering the overall computational cost. 

At the early stages of the ADS process, subdomains can be relatively large, and the accuracy of the HOTM may be insufficient. In such cases, the reference trajectory of the subdomain center may not adequately represent the behavior of the entire region. This can lead to overly conservative decisions. For example, the center trajectory may collide with the Moon, while other trajectories within the same subdomain may remain collision-free. To address this issue, a set of user-defined thresholds for the maximum size of an infeasible subdomain is introduced: $(\Delta x_0^{\max },\Delta \dot x_0^{\max })$ for the planar case and $(\Delta x_0^{\max },\Delta \dot x_0^{\max },\Delta z_0^{\max },\Delta \dot z_0^{\max })$ for the spatial case. If a subdomain is flagged as infeasible but its size exceeds this threshold, it is not immediately discarded but instead further subdivided. The splitting direction is chosen based on the variable that contributes the most to the size excess. For instance, in the planar case, we monitor the current size of the subdomain in both ${x_0}$ and ${\dot x_0}$ directions. If the \emph{k}-th subdomain exceeds the allowable limits in either direction (\emph{e.g.}, $\Delta x_0^{(k)} > \Delta x_0^{\max }$ or $\Delta \dot x_0^{(k)} > \Delta \dot x_0^{\max }$), it is split along the dimension with the largest relative excess. This ensures that large regions are refined appropriately, reducing the risk of prematurely discarding potentially feasible solutions. In addition, if the subdomain is deemed feasible, it is treated as in standard ADS \cite{Wittig2015}: the truncation error of the polynomial is estimated, and if refinement is required, the domain is split along the direction of maximum estimated error. Moreover, following the standard ADS, the maximal number of domain splits, $N_{\max}$, is defined to prevent infinite splitting \cite{Wittig2015}.

Algorithm~\ref{alg2} summarizes the ADS construction procedure for the planar CRTBP case. To improve computational efficiency, \textbf{lines 5} through \textbf{37} (in Algorithm~\ref{alg2}) can be executed in parallel, since each subdomain can be processed independently during feasibility check and HOTM construction. The implementation of ADS for the spatial CRTBP case follows the same procedures, with the only difference being the inclusion of additional state variables in the \emph{z}-direction. For this reason, the spatial case is not detailed separately.

\begin{algorithm}[!h]
\caption{Pseudocode of the implementation of the ADS process in the planar case.}
\label{alg2}
\begin{algorithmic}[1]
\State{Construct the initial domain ${{\mathscr D}_0}$ according to the central point $({x_0},{\dot x_0})$, half-lengths $\Delta {x_0}$ and $\Delta {\dot x_0}$, which is to be addressed by the ADS.} \Comment{using Eq.~\eqref{eq22}}
\State{Define ${\mathscr D}_0^{(1)} = {{\mathscr D}_0}$ to initialize the ADS process: $\mathscr M=\{{\mathscr D}_0^{(1)}\}$.}
\While{subdomains requiring further splitting}
    \For{each subdomain ${\mathscr D}_0^{(k)} \in {\mathscr M}$}
        \State{Isolate $\dot y_0^2$ term according to the given Jacobi constant $C_J$.} \Comment{using Eq.~\eqref{eq28}}
        \If{$\dot y_0^2 < 0$} \Comment{identified the third type of infeasibility}
            \State{Flag ${\mathscr D}_0^{(k)}$ as infeasible.}
        \ElsIf{$\dot y_0^2 \geq 0$}
            \State{Determine the velocity ${\dot y_0}$.} \Comment{using Eq.~\eqref{eq28}}
            \State{Construct the reference state ${\boldsymbol{x}_0} = {[{x_0},0,{\dot x_0},{\dot y_0}]^T}$.}
            \State{Numerically propagate the reference orbit within the interval $[0, \rm{ToF}_{\max}]$.} \Comment{without DA}
            \If{infeasibility 1 or 2 detected} \Comment{identified the first or second type of infeasibility}
                \State{Flag ${\mathscr D}_0^{(k)}$ as infeasible.}
            \ElsIf{feasibility conditions 1 and 2 satisfied}
                \State{Derive the HOTM ${\mathscr T}_{{\rm{HOTM}}}^{(k)}$ of the \emph{k}-th subdomain ${\mathscr D}_0^{(k)}$ (within DA).} \Comment{using Algorithm~\ref{alg1}}
                \State{Determine the overestimated bound of the final return state.} \Comment{using LDB \cite{makino2005range, Armellin2009CMDA}}
                \If{intersect with at least one feasible subdomain in ADS set ${\mathscr M}$}
                    \State{Flag ${\mathscr D}_0^{(k)}$ as feasible.}
                \ElsIf{intersect with no feasible subdomain} \Comment{identified the fourth type of infeasibility}
                    \State{Flag ${\mathscr D}_0^{(k)}$ as infeasible.}
                \EndIf
            \EndIf
        \EndIf
        \If{subdomain ${\mathscr D}_0^{(k)}$ is feasible}
            \State{Estimate the truncation error.} \Comment{following the standard ADS \cite{Wittig2015}}
            \State{Determine whether it needs to be split (according to $\varepsilon$ and $N_{\max}$).}
            \If{further splitting is required}
                \State{Determine the splitting direction and split the subdomain.} \Comment{following the standard ADS \cite{Wittig2015}}
            \EndIf
        \ElsIf{subdomain ${\mathscr D}_0^{(k)}$ is infeasible}
            \If{size of ${\mathscr D}_0^{(k)}$ exceeds the predefined maximum size}
                \State{Split it along the dimension with the largest relative excess.}
            \Else
                \State{Remove it from the ADS set ${\mathscr M}$.}
            \EndIf
        \EndIf
    \EndFor
    \State{Update ADS set ${\mathscr M}$.}
\EndWhile
\State{Return the ADS set ${\mathscr M}$.}
\end{algorithmic}
\end{algorithm}

%%%%%%%%%%%%%%%%%%%%%%%%%%%%%%%%%%%%%%%%%%%%%%%%%%%%%%%%%%%%%%%%%%%%%%%%%%%%%%%%%%%%%%%%%%%
%%%%%%%%%%%%%%%%%%%%%%%%%%%%%%%%%%%%%%%%%%%%%%%%%%%%%%%%%%%%%%%%%%%%%%%%%%%%%%%%%%%%%%%%%%%
\section{Identifying Fixed Points with Polynomial Optimization} \label{sec:Identifying Fixed Points Using Polynomial Optimizatio}
This section presents a two-stage polynomial optimization framework for identifying fixed points across a set of HOTMs defined over multiple subdomains. The method is designed to efficiently locate fixed points (in the Poincaré section) in nonlinear dynamical systems without direct enumeration of all possible trajectories.

To facilitate a clear and concise presentation of the fixed-point identification process, several new symbols representing the states in the Poincaré section are introduced herein. For the planar case, let $\boldsymbol{\bar X}_0^{(k)} = {[x_0^{(k)},\dot x_0^{(k)}]^T}$ and $\delta \boldsymbol{X}_0^{(k)} = {[\delta x_0^{(k)},\delta \dot x_0^{(k)}]^T}$, with $\left| {\delta x_0^{(k)}} \right| \le \Delta x_0^{(k)}$ and $\left| {\delta \dot x_0^{(k)}} \right| \le \Delta \dot x_0^{(k)}$. Then, we define $\boldsymbol{\bar X}_f^{(k)} = {[x_f^{(k)},\dot x_f^{(k)}]^T}$, where $x_f^{(k)}$ and $\dot x_f^{(k)}$ denote the return state propagated based on the central point (of the \emph{k}-th subdomain ${\mathscr D}_0^{(k)}$) along the \emph{x}-axis position and velocity, respectively. Using the derived HOTM, one has
\begin{equation} \label{eq32}
    \delta \boldsymbol{X}_f^{(k)} = \left( {\begin{array}{*{20}{c}}
        {\delta x_f^{(k)}}\\
        {\delta \dot x_f^{(k)}}
    \end{array}} \right) \approx {\mathscr T}_{{\rm{HOTM}}}^{(k)}(\delta \boldsymbol{X}_0^{(k)}) \,.
\end{equation}
The above definitions of the planar case can be easily extended to the spatial case, \emph{i.e.}, $\boldsymbol{\bar X}_0^{(k)} = {[x_0^{(k)},\dot x_0^{(k)},z_0^{(k)},\dot z_0^{(k)}]^T}$, $\boldsymbol{\bar X}_f^{(k)} = {[x_f^{(k)},\dot x_f^{(k)},z_f^{(k)},\dot z_f^{(k)}]^T}$, $\delta \boldsymbol{X}_0^{(k)} = {[\delta x_0^{(k)},\delta \dot x_0^{(k)},\delta z_0^{(k)},\delta \dot z_0^{(k)}]^T}$, and $\delta \boldsymbol{X}_f^{(k)} = {[\delta x_f^{(k)},\delta \dot x_f^{(k)},\delta z_f^{(k)},\delta \dot z_f^{(k)}]^T}$. With these newly defined variables, the initial and final return state of a neighboring trajectory can be represented as
\begin{equation} \label{eq33}
    \boldsymbol{X}_0^{(k)} = \boldsymbol{\bar X}_0^{(k)} + \delta \boldsymbol{X}_0^{(k)} \in {\mathscr D}_0^{(k)}
\end{equation}
and
\begin{equation} \label{eq34}
    \boldsymbol{X}_f^{(k)} = \boldsymbol{\bar X}_f^{(k)} + \delta \boldsymbol{X}_f^{(k)} \approx \boldsymbol{\bar X}_f^{(k)} + {\mathscr T}_{{\rm{HOTM}}}^{(k)}(\delta \boldsymbol{X}_0^{(k)}) \,,
\end{equation}
respectively.

%%%%%%%%%%%%%%%%%%%%%%%%%%%%%%%%%%%%%%%%%%%%%%%%%%%%%%%%%%%%%%%%%%%%%%%%%%%%%%%%%%%%%%%%%%%
\subsection{Identification of Combinable HOTM Sequences} \label{sec:Identification of Combinable HOTM Sequences}
The first stage of the proposed method focuses on the identification of combinable subdomain sequences that can potentially form continuous ballistic trajectories when connected through successive HOTMs. 
This step is crucial because, in practice, the number of subdomains obtained in the ADS process is often large, making a brute-force search for fixed points across all possible combinations computationally intractable. Therefore, before attempting to solve for fixed points directly, it is essential to determine which subdomains (and the corresponding HOTMs) are mutually compatible and can be meaningfully combined. The goal of this first stage is to perform a fast and rigorous pruning of infeasible connections, thereby narrowing down the search space for the subsequent optimization stage.

\begin{figure}[!h]
    \centering
    \includegraphics[width=0.75\linewidth]{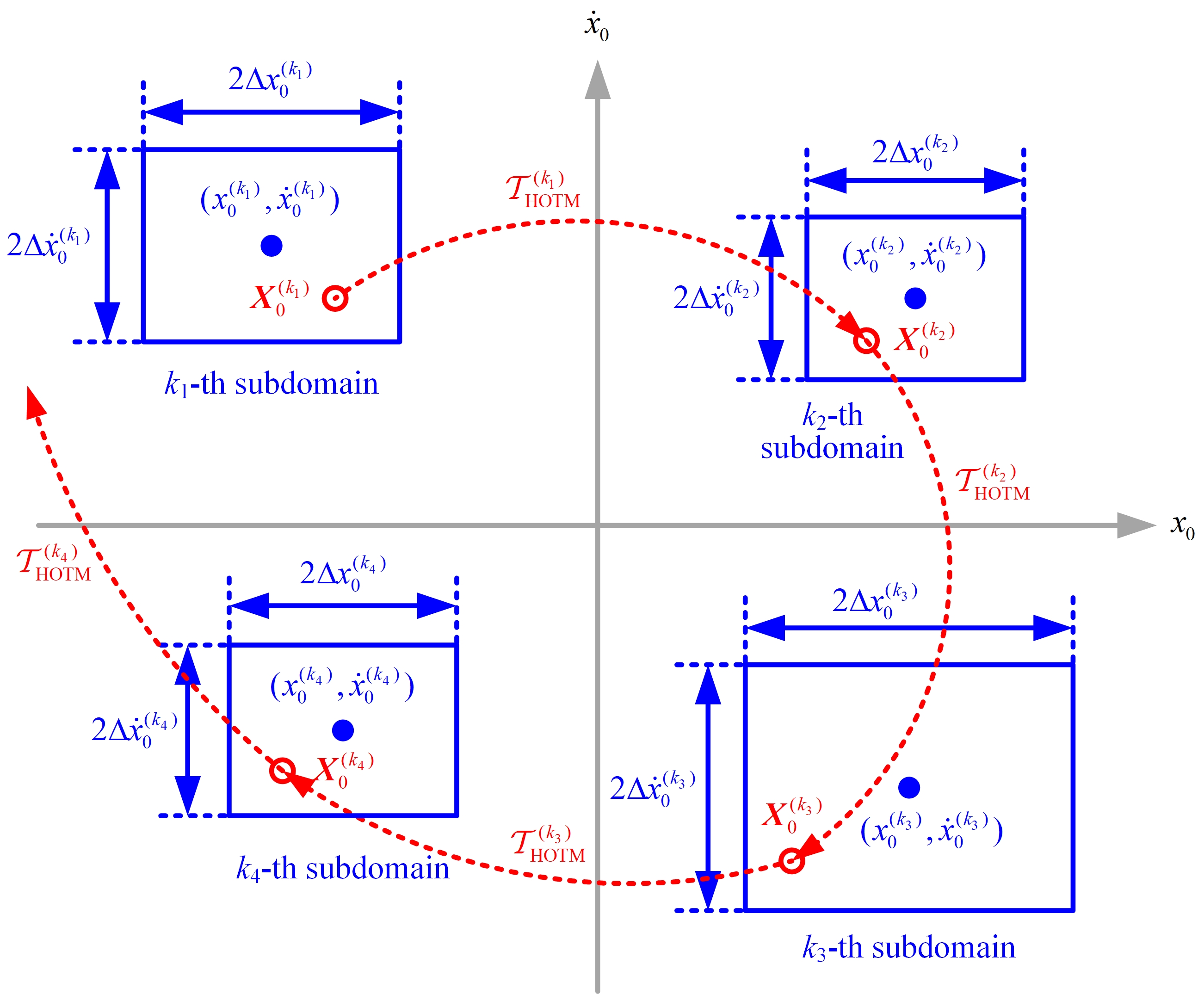}
    \caption{\label{fig2} An illustration of the combinability condition in the planar case.}
\end{figure}

Figure~\ref{fig2} illustrates the combinability condition for a set of four subdomains ($k_1$, $k_2$, $k_3$, and $k_4$) and their associated HOTMs in the planar case. Given an ordered sequence of subdomains
\begin{equation} \label{eq35}
    \left\{ {{\mathscr D}_0^{({k_1})},{\mathscr D}_0^{({k_2})}, \cdots ,{\mathscr D}_0^{({k_i})}, \cdots ,{\mathscr D}_0^{({k_n})}} \right\} \,,
\end{equation}
where $n$ is the total length of the subdomain sequence, only adjacent pairs that are consecutively connected are subject to continuity constraints. In other words, the output region of the HOTM from the $k_i$-th subdomain must intersect the input domain of the $k_{i+1}$-th subdomain for the combination to be considered feasible. No such constraint is imposed between the first and last subdomains in the sequence, as the overall path is not required to be cyclic at this stage. 

Therefore, in general, the combinability condition for the subdomain sequence in Eq.~\eqref{eq35} is expressed as
\begin{equation} \label{eq36}
    \begin{array}{*{20}{c}}
        {\boldsymbol{X}_0^{({k_{i + 1}})} = \boldsymbol{X}_f^{({k_i})},}&{i \in \{ 1,2, \cdots ,n - 1\} }
    \end{array} \,.
\end{equation}
Substitution of Eqs.~\eqref{eq33}-\eqref{eq34} into Eq.~\eqref{eq36} yields
\begin{equation} \label{eq37}
    \boldsymbol{\bar X}_0^{({k_{i + 1}})} + \delta \boldsymbol{X}_0^{({k_{i + 1}})} \approx \boldsymbol{\bar X}_f^{({k_i})} + {\mathscr T}_{{\rm{HOTM}}}^{({k_i})}(\delta \boldsymbol{X}_0^{({k_i})}) \,.
\end{equation}

Since the combinability condition described above involves solving a set of polynomial equality constraints, which is generally difficult to handle directly, we reformulate the problem as a polynomial optimization problem (POP) by minimizing the residual between the two sides of the condition:
\begin{equation} \label{eq38}
    \begin{array}{*{20}{l}}
        {{\rm{find}}}&{\delta \boldsymbol{X}_0^{({k_i})}}\\
        {\min }&{J = \underbrace {\sum\nolimits_{i = 1}^{n - 1} {{{\left\| {\boldsymbol{X}_f^{({k_i})} - \boldsymbol{X}_0^{({k_{i + 1}})}} \right\|}^2}} }_{{\rm{Continuity\,condition}}}}\\
        {{\rm{s}}{\rm{.t}}{\rm{.}}}&{\boldsymbol{X}_0^{({k_i})} = \boldsymbol{\bar X}_0^{({k_i})} + \delta \boldsymbol{X}_0^{({k_i})} \in {\mathscr D}_0^{({k_i})}}\\
        {}&{\boldsymbol{X}_f^{({k_i})} = \boldsymbol{\bar X}_f^{({k_i})} + {\mathscr T}_{{\rm{HOTM}}}^{({k_i})}(\delta \boldsymbol{X}_0^{({k_i})})}
    \end{array} \,.
\end{equation}
Here, the approximate equality symbol is replaced with a strict equality sign in the objective function $J$, since the approximation error of the HOTM is not explicitly accounted for during the optimization process.

In this work, the POP in Eq.~\eqref{eq38} is referred to as the first-stage POP. It is solved using a recursive polynomial optimization (RPO) method, which will be introduced in Sec.~\ref{sec:Recursive Polynomial Optimizations}. If the minimal objective $J$ is lower than a given threshold ($\varepsilon_1=10^{-9}$ in this work), the $n$ selected HOTMs from ${\mathscr T}_{{\rm{HOTM}}}^{({k_1})}$ to ${\mathscr T}_{{\rm{HOTM}}}^{({k_n})}$ are considered as combinable, indicating that there exists at least one ballistic (natural) trajectory that can connect the $n$ selected subdomains in Eq.~\eqref{eq35}. Compared to the LDB-based overestimation approach (employed in the ADS process), this POP-based method offers higher accuracy in evaluating combinability. On the other hand, it is computationally more expensive. Therefore, it is not employed during the ADS construction stage, where efficiency is prioritized.

To evaluate whether a sequence of $n=2$ subdomains (and HOTMs) can be combined, all $K \times K$ possible pairs must be tested, where $K$ is the number of feasible subdomains. However, for $n \geq 3$, it is not necessary to enumerate all $K^n$ possible combinations directly. Instead, the combinability of an \emph{n}-length sequence can be recursively determined based on the combinability of (\emph{n}-1)-length sequences. Specifically, a sequence $({k_1},{k_2}, \cdots ,{k_{n - 1}},{k_n})$ is combinable if both the subsequence $({k_1},{k_2}, \cdots ,{k_{n - 1}})$ is known to be combinable and the pair $({k_{n - 1}},{k_n})$ is also combinable. This recursive property significantly reduces the computational burden and allows for the efficient construction of combinable subdomain (and HOTM) chains of arbitrary length.

%%%%%%%%%%%%%%%%%%%%%%%%%%%%%%%%%%%%%%%%%%%%%%%%%%%%%%%%%%%%%%%%%%%%%%%%%%%%%%%%%%%%%%%%%%%
\subsection{Fixed-Point Identification Based on Combinable HOTMs} \label{sec:Fixed-Point Identification Based on Combinable HOTMs}
This subsection presents the second stage of the proposed method, which aims to identify fixed points (POs) based on the combinable HOTM sequences obtained in the first stage. After determining feasible subdomain sequences that can be connected through the HOTMs, additional fixed-point constraints are introduced to locate fixed-point solutions that return exactly to their initial state after completing a full sequence. The resulting method enables the efficient search for POs in both planar and spatial CRTBP.

\begin{figure}[!h]
    \centering
    \includegraphics[width=0.75\linewidth]{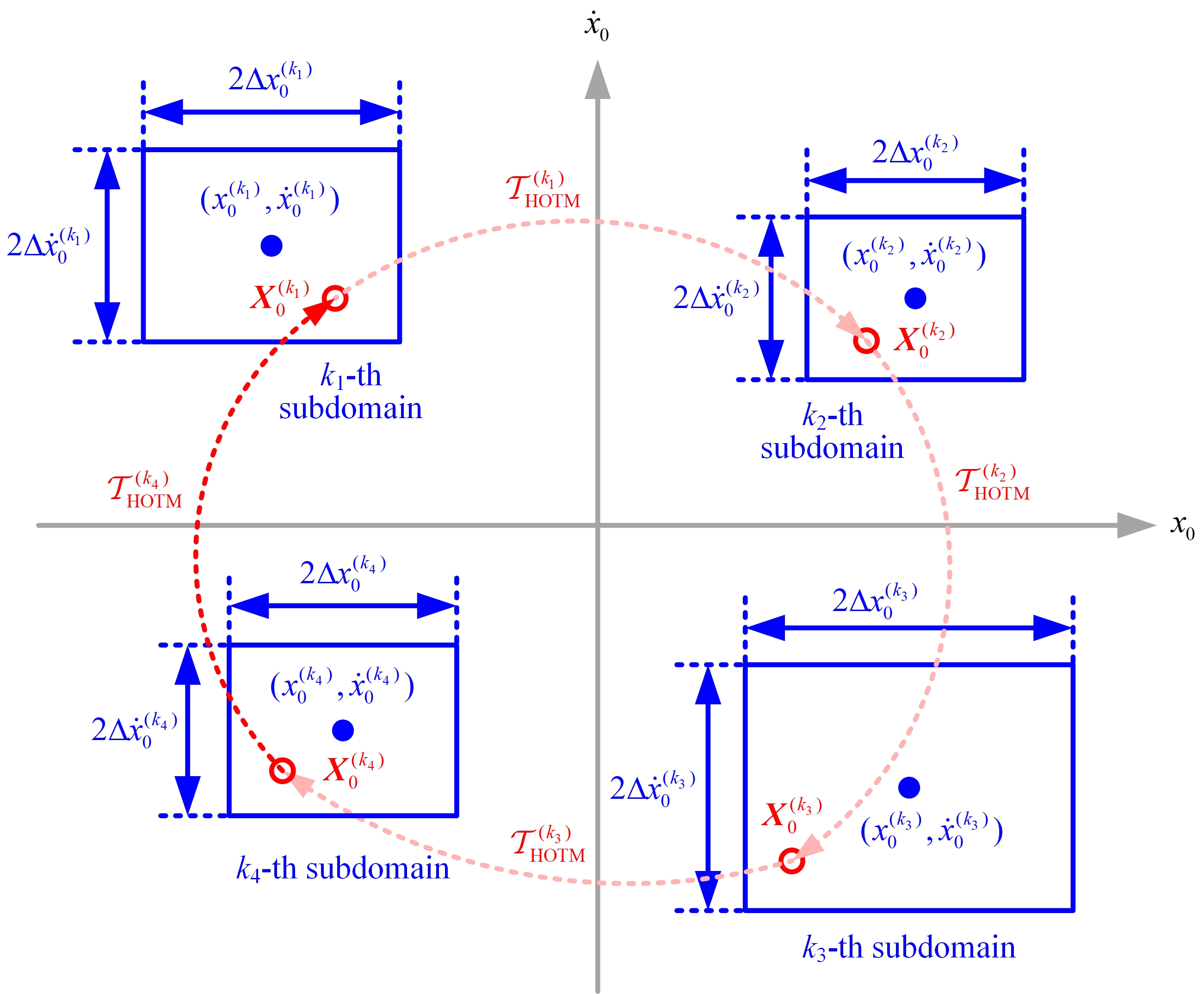}
    \caption{\label{fig3} An illustration of the fixed point conditions in the planar case.}
\end{figure}

Figure~\ref{fig3} illustrates the concept of fixed-point identification in the planar case ($n=4$). Compared to Fig.~\ref{fig2}, the key difference lies in the additional periodicity constraint, represented by the red dashed line connecting the final state of the last HOTM  back to the initial state of the first subdomain, namely $\boldsymbol{\bar X}_f^{({k_4})} + {\mathscr T}_{{\rm{HOTM}}}^{({k_4})}(\delta \boldsymbol{X}_0^{({k_4})}) \approx \boldsymbol{X}_0^{({k_1})}$. 
The continuity conditions between successive HOTMs (shown in light red) remain unchanged from Fig.~\ref{fig2}, ensuring that the sequence forms a valid concatenation of subdomains. The periodic constraint enforces that the entire trajectory is closed, forming a PO described by a fixed point in the Poincaré section:
\begin{equation} \label{eq39}
    \boldsymbol{\bar X}_0^{({k_1})} + \delta \boldsymbol{X}_0^{({k_1})} \approx \boldsymbol{\bar X}_f^{({k_n})} + {\mathscr T}_{{\rm{HOTM}}}^{({k_n})}(\delta \boldsymbol{X}_0^{({k_n})}) \,.
\end{equation}

With the additional periodicity constraint, a POP for identifying \emph{n}-revolution fixed points is formulated as
\begin{equation} \label{eq40}
    \begin{array}{*{20}{l}}
        {{\rm{find}}}&{\delta \boldsymbol{X}_0^{({k_i})}}\\
        {\min }&{J = \underbrace {\sum\nolimits_{i = 1}^{n - 1} {{{\left\| {\boldsymbol{X}_f^{({k_i})} - \boldsymbol{X}_0^{({k_{i + 1}})}} \right\|}^2}} }_{{\rm{Continuity\,condition}}} + \underbrace {{{\left\| {\boldsymbol{X}_0^{({k_1})} - \boldsymbol{X}_f^{({k_n})}} \right\|}^2}}_{{\rm{Periodicity\,constraint}}}}\\
        {{\rm{s}}{\rm{.t}}{\rm{.}}}&{\boldsymbol{X}_0^{({k_i})} = \boldsymbol{\bar X}_0^{({k_i})} + \delta \boldsymbol{X}_0^{({k_i})} \in {\mathscr D}_0^{({k_i})}}\\
        {}&{\boldsymbol{X}_f^{({k_i})} = \boldsymbol{\bar X}_f^{({k_i})} + {\mathscr T}_{{\rm{HOTM}}}^{({k_i})}(\delta \boldsymbol{X}_0^{({k_i})})}
    \end{array} \,,
\end{equation}
where the first and second terms in the objective function $J$ correspond to the continuity condition in Eq.~\eqref{eq36} and the periodicity constraint in Eq.~\eqref{eq39}. The POP in Eq.~\eqref{eq40} is referred to as second-stage POP, which is also solved using the RPO method. Like the descriptions in Sec.~\ref{sec:Identification of Combinable HOTM Sequences}, a solution is considered as a valid fixed point when the minimal objective $J$ of the POP in Eq.~\eqref{eq40} is lower than a given threshold ($\varepsilon_2=10^{-6}$ in this work).

One special case for the POP in Eq.~\eqref{eq44} is when $n=1$:
\begin{equation} \label{eq41}
    \begin{array}{*{20}{l}}
        {{\rm{find}}}&{\delta \boldsymbol{X}_0^{({k_1})}}\\
        {\min }&{J = \underbrace {{{\left\| {\boldsymbol{X}_f^{({k_1})} - \boldsymbol{X}_0^{({k_1})}} \right\|}^2}}_{{\rm{Periodicity\,constraint}}}}\\
        {{\rm{s}}{\rm{.t}}{\rm{.}}}&{\boldsymbol{X}_0^{({k_1})} = \boldsymbol{\bar X}_0^{({k_1})} + \delta \boldsymbol{X}_0^{({k_1})} \in {\mathscr D}_0^{({k_1})}}\\
        {}&{\boldsymbol{X}_f^{({k_1})} = \boldsymbol{\bar X}_f^{({k_1})} + {\mathscr T}_{{\rm{HOTM}}}^{({k_1})}(\delta \boldsymbol{X}_0^{({k_1})})}
    \end{array} \,,
\end{equation}
where the continuity condition doesn’t exist (as only one single subdomain is considered). The solutions of the POP in Eq.~\eqref{eq41} correspond to single-revolution POs. Representative examples are Lyapunov orbits and DROs.

Although one might consider directly enumerating all possible sequences and solving the second-stage POP in Eq.~\eqref{eq40} to obtain fixed points, this approach entails a significantly higher computational burden. The purpose of the first-stage POP in Eq.~\eqref{eq38} is to identify all combinable subdomain (and HOTM) sequences efficiently. Once these combinable sequences of length \emph{n} are known, feasible sequences of length \emph{n}+1 can be easily constructed based on the existing results, thus avoiding most unnecessary computations. If one skips this step and directly solves the second-stage POP for all possible sequences, the total number of optimization problems to solve grows exponentially as $K^n$, where \emph{K} is the total number of feasible subdomains. Therefore, the first-stage POP serves as a crucial pruning mechanism to dramatically reduce the computational burden and enable scalable search for multiple-revolution fixed points.

%%%%%%%%%%%%%%%%%%%%%%%%%%%%%%%%%%%%%%%%%%%%%%%%%%%%%%%%%%%%%%%%%%%%%%%%%%%%%%%%%%%%%%%%%%%
\subsection{Recursive Polynomial Optimization} \label{sec:Recursive Polynomial Optimizations}
Both POPs introduced in Sec.~\ref{sec:Identification of Combinable HOTM Sequences} and Sec.~\ref{sec:Fixed-Point Identification Based on Combinable HOTMs} are solved using the RPO method. The RPO method is not a newly developed approach. It is a nonlinear optimization framework that combines polynomial approximation with convex programming to efficiently solve problems involving polynomial objectives and constraints. Originally proposed by Pavanello \emph{et al.}~\cite{Pavanello2024IEEE} for collision avoidance applications, the classical RPO framework incrementally increases the polynomial order and solves a sequence of convex subproblems, thereby improving the approximation quality with each iteration. This strategy enables robust handling of nonlinearities while ensuring convergence even in challenging scenarios. Zhou \emph{et al.}~\cite{zhou2025maneuver} later introduced an improved variant tailored for maneuver detection. Instead of gradually increasing the approximation order, their approach directly utilizes the full high-order polynomial expansion and applies recursive linearization around the current estimate. This significantly reduces computational cost and the number of required iterations, while retaining the accuracy and robustness of the solution.

In this work, the improved RPO scheme proposed by Zhou \emph{et al.}~\cite{zhou2025maneuver} is employed to efficiently solve the two POPs arising from HOTM combinability and fixed-point constraints, respectively. Although the RPO method is not new, its applicability is limited to problems where both the objective function and constraints are polynomial in nature. Any additional non-polynomial terms must be reformulated or approximated in a convex-compatible manner. In other words, RPO is only responsible for handling the polynomial components of the optimization problem, and it does not inherently address general nonlinearities or complex constraint types. As a result, additional reformulation is required to cast the previously defined POPs into a form compatible with RPO. This includes restructuring the objective and constraint expressions to isolate or eliminate any non-polynomial elements.

The only non-polynomial component of both POPs lies in the objective function (note that the constraint $\boldsymbol{X}_0^{({k_i})} = \boldsymbol{\bar X}_0^{({k_i})} + \delta \boldsymbol{X}_0^{({k_i})} \in {\mathscr D}_0^{({k_i})}$ is linear). Take the second-stage POP as an example, to reformulate the objective function, two types of auxiliary variables $s_c$ and $s_p$ are defined as
\begin{equation} \label{eq42}
    \begin{array}{*{20}{c}}
    {{{\left\| {\boldsymbol{X}_f^{({k_i})} - \boldsymbol{X}_0^{({k_{i + 1}})}} \right\|}^2} \le {s_{c,i}},}&{i \in \{ 1,2, \cdots ,n - 1\} }
    \end{array} \,,
\end{equation}
\begin{equation} \label{eq43}
    {\left\| {\boldsymbol{X}_0^{({k_1})} - \boldsymbol{X}_f^{({k_n})}} \right\|^2} \le {s_p} \,.
\end{equation}
Substituting Eqs.~\eqref{eq42}-\eqref{eq43} into the POP in Eq.~\eqref{eq44}, one has
\begin{equation} \label{eq44}
    \begin{array}{*{20}{l}}
        {{\rm{find}}}&{\delta \boldsymbol{X}_0^{({k_i})}}\\
        {\min }&{J = \underbrace {\sum\nolimits_{i = 1}^{n - 1} {{s_{c,i}}} }_{{\rm{Continuity\,condition}}} + \underbrace {{s_p}}_{{\rm{Periodicity\,constraint}}}}\\
        {{\rm{s}}{\rm{.t}}{\rm{.}}}&{\boldsymbol{X}_0^{({k_i})} = \boldsymbol{\bar X}_0^{({k_i})} + \delta \boldsymbol{X}_0^{({k_i})} \in {\mathscr D}_0^{({k_i})}}\\
        {}&{\boldsymbol{X}_f^{({k_i})} = \boldsymbol{\bar X}_f^{({k_i})} + {\mathscr T}_{{\rm{HOTM}}}^{({k_i})}(\delta \boldsymbol{X}_0^{({k_i})})}\\
        {}&{{{\left\| {\boldsymbol{X}_f^{({k_i})} - \boldsymbol{X}_0^{({k_{i + 1}})}} \right\|}^2} \le {s_{c,i}}}\\
        {}&{{{\left\| {\boldsymbol{X}_0^{({k_1})} - \boldsymbol{X}_f^{({k_n})}} \right\|}^2} \le {s_p}}
    \end{array} \,,
\end{equation}
which is compatible with the RPO method. In addition, the first-stage POP can be reconstructed using the same method as in Eqs.~\eqref{eq42}-\eqref{eq44}.

Following the RPO, given an initial guess for $\delta \boldsymbol{X}_0^{({k_i})}$, labeled as $\delta \boldsymbol{\tilde X}_0^{({k_i})}$ (note that $\delta \boldsymbol{\tilde X}_0^{({k_i})}$ is a numerical vector rather than a variable), one can linearize the HOTM ${\mathscr T}_{{\rm{HOTM}}}^{({k_i})}(\delta \boldsymbol{X}_0^{({k_i})})$ as
\begin{equation} \label{eq46}
    {\mathscr T}_{{\rm{HOTM}}}^{({k_i})}(\delta \boldsymbol{X}_0^{({k_i})}) \approx {\mathscr A}_{{\rm{HOTM}}}^{({k_i})}\Delta \boldsymbol{X}_0^{({k_i})} + {\mathscr B}_{{\rm{HOTM}}}^{({k_i})} \,,
\end{equation}
where $\Delta \boldsymbol{X}_0^{({k_i})} = \delta \boldsymbol{X}_0^{({k_i})} - \delta \boldsymbol{\tilde X}_0^{({k_i})}$, ${\mathscr A}_{{\rm{HOTM}}}^{({k_i})} = {\nabla {\mathscr T}_{{\rm{HOTM}}}^{({k_i})}(\delta \boldsymbol{\tilde  X}_0^{({k_i})})}$ and ${\mathscr B}_{{\rm{HOTM}}}^{({k_i})} = {\mathscr T}_{{\rm{HOTM}}}^{({k_i})}(\delta \boldsymbol{\tilde X}_0^{({k_i})})$.

Then, a subproblem of the POP in Eq.~\eqref{eq44} is given as
\begin{equation} \label{eq49}
    \begin{array}{*{20}{l}}
        {{\rm{find}}}&{\Delta \boldsymbol{X}_0^{({k_i})}}\\
        {\min }&{J = \underbrace {\sum\nolimits_{i = 1}^{n - 1} {{s_{c,i}}} }_{{\rm{Continuity\,condition}}} + \underbrace {{s_p}}_{{\rm{Periodicity\,constraint}}}}\\
        {{\rm{s}}{\rm{.t}}{\rm{.}}}&{\boldsymbol{X}_0^{({k_i})} = \boldsymbol{\bar X}_0^{({k_i})} + \delta \boldsymbol{\tilde X}_0^{({k_i})} + \Delta \boldsymbol{X}_0^{({k_i})} \in {\mathscr D}_0^{({k_i})}}\\
        {}&{\boldsymbol{X}_f^{({k_i})} = \boldsymbol{\bar X}_f^{({k_i})} + {\mathscr A}_{{\rm{HOTM}}}^{({k_i})}\Delta \boldsymbol{X}_0^{({k_i})} + {\mathscr B}_{{\rm{HOTM}}}^{({k_i})}}\\
        {}&{{{\left\| {\boldsymbol{X}_f^{({k_i})} - \boldsymbol{X}_0^{({k_{i + 1}})}} \right\|}^2} \le {s_{c,i}}}\\
        {}&{{{\left\| {\boldsymbol{X}_0^{({k_1})} - \boldsymbol{X}_f^{({k_n})}} \right\|}^2} \le {s_p}}
    \end{array} \,,
\end{equation}
which becomes a second-order cone optimization problem (SCOP) and can be easily solved using convex optimizers like MOSEK \footnote{Available at \url{https://www.mosek.com/}.}. Once the SCOP is solved, one can update the guess as $\delta \boldsymbol{\tilde X}_0^{({k_i})} \leftarrow \delta \boldsymbol{\tilde X}_0^{({k_i})} + \Delta \boldsymbol{X}_0^{({k_i})}$ until a predefined threshold $\eta$ is satisfied: $\left\| {\Delta \boldsymbol{X}_0^{({k_i})}} \right\| \le \eta $. For the first iteration, the initial guess can be directly set as a zero vector: $\delta \boldsymbol{\tilde X}_0^{({k_i})} = {\mathbf{0}_2}$ (planar case) or $\delta \boldsymbol{\tilde X}_0^{({k_i})} = {\mathbf{0}_4}$ (spatial case). Similarly, the simpler first-stage POP in Eq.~\eqref{eq38} can be solved by the RPO with the same procedures shown in Eqs.~\eqref{eq46}-\eqref{eq49}.

%%%%%%%%%%%%%%%%%%%%%%%%%%%%%%%%%%%%%%%%%%%%%%%%%%%%%%%%%%%%%%%%%%%%%%%%%%%%%%%%%%%%%%%%%%%
\subsection{Overall Procedure} \label{sec:Overall Procedure}
Based on the descriptions in Sec.~\ref{sec:High-Order Transfer Map}, Sec.~\ref{sec:Identification of Combinable HOTM Sequences}, Sec.~\ref{sec:Fixed-Point Identification Based on Combinable HOTMs}, and Sec.~\ref{sec:Recursive Polynomial Optimizations}, the overall procedure of the proposed method for identifying fixed points is summarized in Algorithm~\ref{alg3}. The proposed method constructs single-return HOTMs and identifies fixed points by optimizing over concatenated sequences of subdomains. Since the HOTMs are designed to map trajectories from a Poincaré section back to itself within one revolution, longer-period fixed points can be found by chaining multiple HOTMs without the need to reconstruct them for each revolution count. This makes the proposed framework highly scalable and theoretically capable of identifying fixed points with arbitrary revolution numbers using a single set of precomputed HOTMs. 

\begin{algorithm}[!h]
\caption{Pseudocode for identifying \emph{n}-revolution fixed points.} \label{alg3}
\begin{algorithmic}[1]
\State{Define the initial domain and derive the HOTMs within the ADS framework.} \Comment{using Algorithms~\ref{alg1} and~\ref{alg2}}
\State{Recursively determine all \emph{n}-length candidate sequences based on the (\emph{n}-1)-length sequences.}
\For{each \emph{n}-length candidate sequence}
    \State{Solve the first-stage POP using the RPO to obtain the minimal $J^*$.} \Comment{the POP in Eq.~\eqref{eq38}}
    \If{$J^*\leq\varepsilon_1$}
        \State{The sequence is considered combinable.} \Comment{preserved for determining (\emph{n}+1)-length combinable sequence}
    \EndIf
\EndFor
\For{each \emph{n}-length combinable sequence}
    \State{Solve the second-stage POP using the RPO to obtain the minimal $J^*$.} \Comment{the POP in Eq.~\eqref{eq44}}
    \If{$J^*\leq\varepsilon_2$}
        \State{The solution of the second-stage POP is added as an \emph{n}-revolution fixed point.}
    \EndIf
\EndFor
\State{Output all \emph{n}-revolution fixed points.}
\end{algorithmic}
\end{algorithm}

The only potential challenge lies in the accumulation of polynomial approximation errors: as the number of revolutions increases, truncation errors may compound and degrade the accuracy of the solution. To address this, the truncation error threshold $\varepsilon$ in Algorithm~\ref{alg2} in the ADS process can be tightened to improve the accuracy of each HOTM, albeit at an increased computational cost. This trade-off allows users to flexibly balance accuracy and efficiency according to specific application needs.

%%%%%%%%%%%%%%%%%%%%%%%%%%%%%%%%%%%%%%%%%%%%%%%%%%%%%%%%%%%%%%%%%%%%%%%%%%%%%%%%%%%%%%%%%%%
%%%%%%%%%%%%%%%%%%%%%%%%%%%%%%%%%%%%%%%%%%%%%%%%%%%%%%%%%%%%%%%%%%%%%%%%%%%%%%%%%%%%%%%%%%%
\section{Numerical Examples} \label{sec:Numerical Examples}
Fixed points (POs) up to the 9-revolution in the Earth-Moon system are identified in this section, using the proposed method. The implementation is developed in Python 3.10 within the PyCharm integrated development environment.

%%%%%%%%%%%%%%%%%%%%%%%%%%%%%%%%%%%%%%%%%%%%%%%%%%%%%%%%%%%%%%%%%%%%%%%%%%%%%%%%%%%%%%%%%%%
\subsection{Parameters of the Earth-Moon CRTBP} \label{sec:Parameters of the Earth-Moon CRTBPs}
The Earth-Moon system is selected for numerical demonstrations as it represents one of the most extensively studied configurations in the CRTBP. The gravitational constant of the Earth-Moon system is given in Table~\ref{tab4}. The following two subsections present the results of fixed-point identification for the planar and spatial cases, respectively. In addition, unless otherwise specified, the nondimensional (nd) unit system is adopted throughout.

\begin{table}[!h]
    \caption{\label{tab4}Parameters of the Earth-Moon system}
    \centering
    \begin{tabular}{llll}
        \hline\hline
        \multicolumn{2}{l}{Parameter} & Symbol & Value \\ \hline
        \multirow{3}{*}{\makecell[l]{Gravitational\\constant}} & Earth & $\mu_E$ & 398600.43543609598 $\rm{km}^3/\rm{s}2$ \\ 
        & Moon & $\mu_M$ & 4902.8000661637961 $\rm{km}^3/\rm{s}2$ \\
        & Earth-Moon system & $\mu$ & 0.012150584269940354 \\ \hline
        \multirow{3}{*}{\makecell[l]{Nondimensional\\unit}} & Length unit & LU & 384400.0 km \\
        & Velocity unit & VU & 1.02454629434750 km/s \\
        & Time unit & TU & 375190.464423878 s \\
        \hline\hline
    \end{tabular}
\end{table}

%%%%%%%%%%%%%%%%%%%%%%%%%%%%%%%%%%%%%%%%%%%%%%%%%%%%%%%%%%%%%%%%%%%%%%%%%%%%%%%%%%%%%%%%%%%
\subsection{Planar Case Results} \label{sec:Planar Case Results}
In the planar case analysis, two distinct scenarios are investigated. The first case focuses on POs in the vicinity of the Moon, leading to the identification of the known P\emph{n}DRO families and the discovery of a previously undocumented family of POs, referred to as P\emph{n}g'. The second case extends the search domain by enlarging the range of $x_0$ and increasing the maximum allowable time-of-flight. This expanded search reveals a number of unusual trajectories that shuttle between both sides of the Earth, resembling in geometry the Earth-Moon cycler orbits.

\subsubsection{First Planar Case} \label{sec:First Planar Case}
The central point of the first planar case is selected as ${x_0} = 0.87$ and ${\dot x_0} = 0$, with the Jacobi constant being $C_J=3.00022$. The half-lengths (of the initial domain ${{\mathscr D}_0}$) along the ${x_0}$ and ${\dot x_0}$ directions are set as $\Delta {x_0} = 0.1$ and $\Delta {\dot x_0} = 0.6$; thus, the initial domain ${{\mathscr D}_0}$ in Eq.~\eqref{eq24} is expressed as
\begin{equation} \label{eq50}
    {{\mathscr D}_0} = [0.77,0.97] \times [ - 0.6,0.6] \,.
\end{equation}
The maximal order of the polynomial expansions is set as 5, and the implementation-specific parameter settings of this planar case are provided in Table~\ref{tab5}. 
These parameter settings are motivated by the fact that the resulting initial domain and the maximum allowable ToF not only cover the DRO corresponding to the given Jacobi constant, but also provide a sufficient margin. 
In particular, the maximum allowable ToF is set to approximately six times the DRO period (approximately 1.5 TU), and the initial domain is chosen also to include the known P3DRO and P4DRO families, thereby ensuring that there are at least three fixed points within the region.

\begin{table}[!h]
    \caption{\label{tab5} User-defined parameters of the first planar case}
    \centering
    \begin{tabular}{lll}
    \hline\hline
        Parameter & Symbol & Value \\ \hline
        Maximum allowable $\rm{ToF}$ & $\rm{ToF}_{\max}$ & 9 TU \\ \hline
        \multirow{2}{*}{\makecell[l]{Minimum allowable\\distance}} & $d_E^{\min }$ & $10^{-3}$ LU \\
        & $d_M^{\min }$ & $10^{-3}$ LU \\ \hline
        \multirow{2}{*}{\makecell[l]{Maximal size of\\infeasible subdomains}} & $\Delta x_0^{\max }$ & $10^{-3}$ LU \\
        & $\Delta \dot x_0^{\max }$ & $10^{-3}$ VU \\ \hline
        \multirow{2}{*}{ADS parameters} & $\varepsilon$ & $10^{-5}$ \\
        & $N_{\max}$ & 30 \\ \hline
        \multirow{3}{*}{\makecell[l]{Two-stage optimization\\thresholds}}  & $\varepsilon_1$ & $10^{-9}$ \\
        & $\varepsilon_2$ & $10^{-6}$ \\
        & $\eta$ & $10^{-6}$ \\ 
    \hline\hline
    \end{tabular}
\end{table}

\begin{figure}[!h]
    \centering
    \subfigure[Subdomains]{
        \label{fig4a}
        \includegraphics[width=0.48\linewidth]{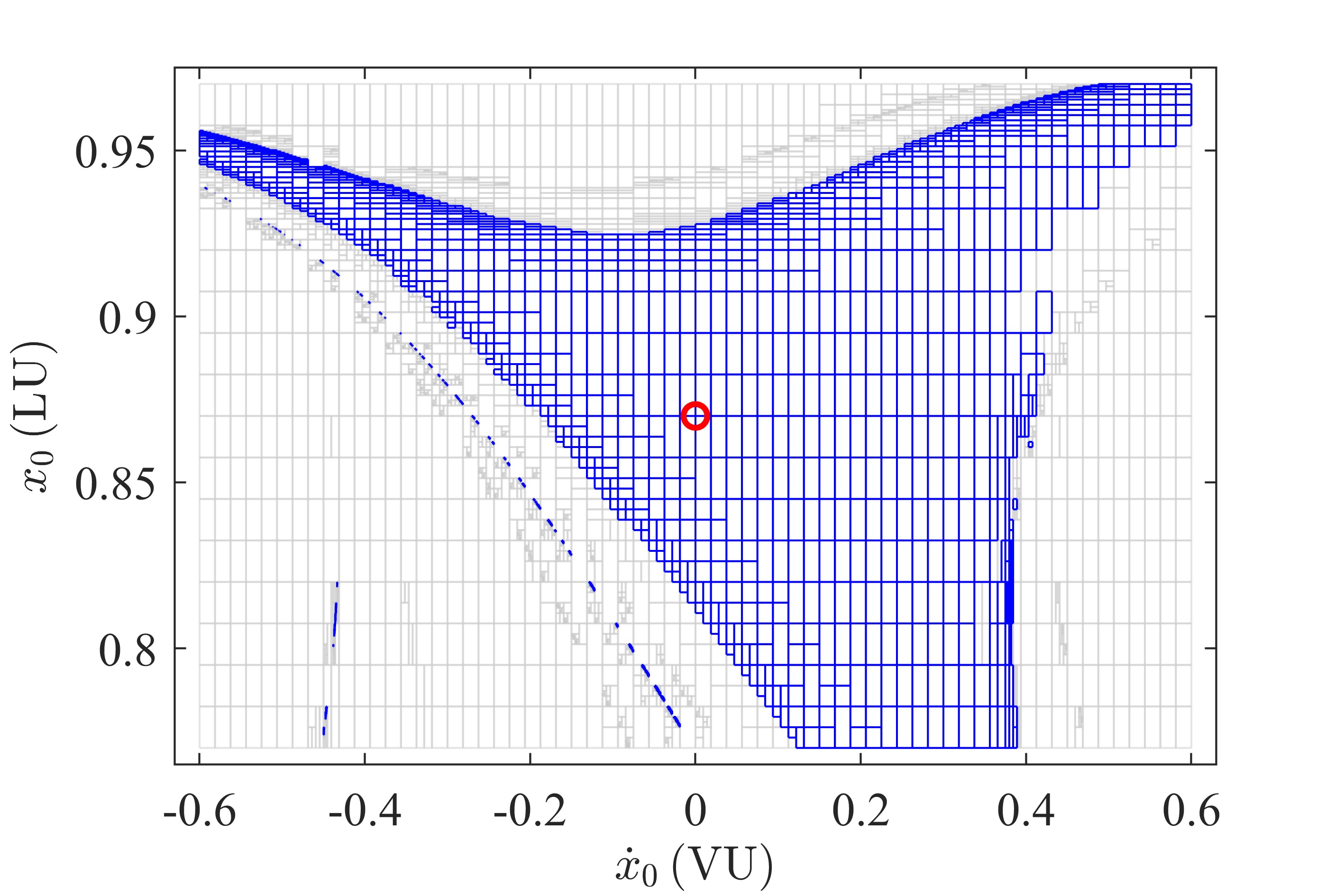}
    }
    \subfigure[ToF]{
        \label{fig4b}
        \includegraphics[width=0.48\linewidth]{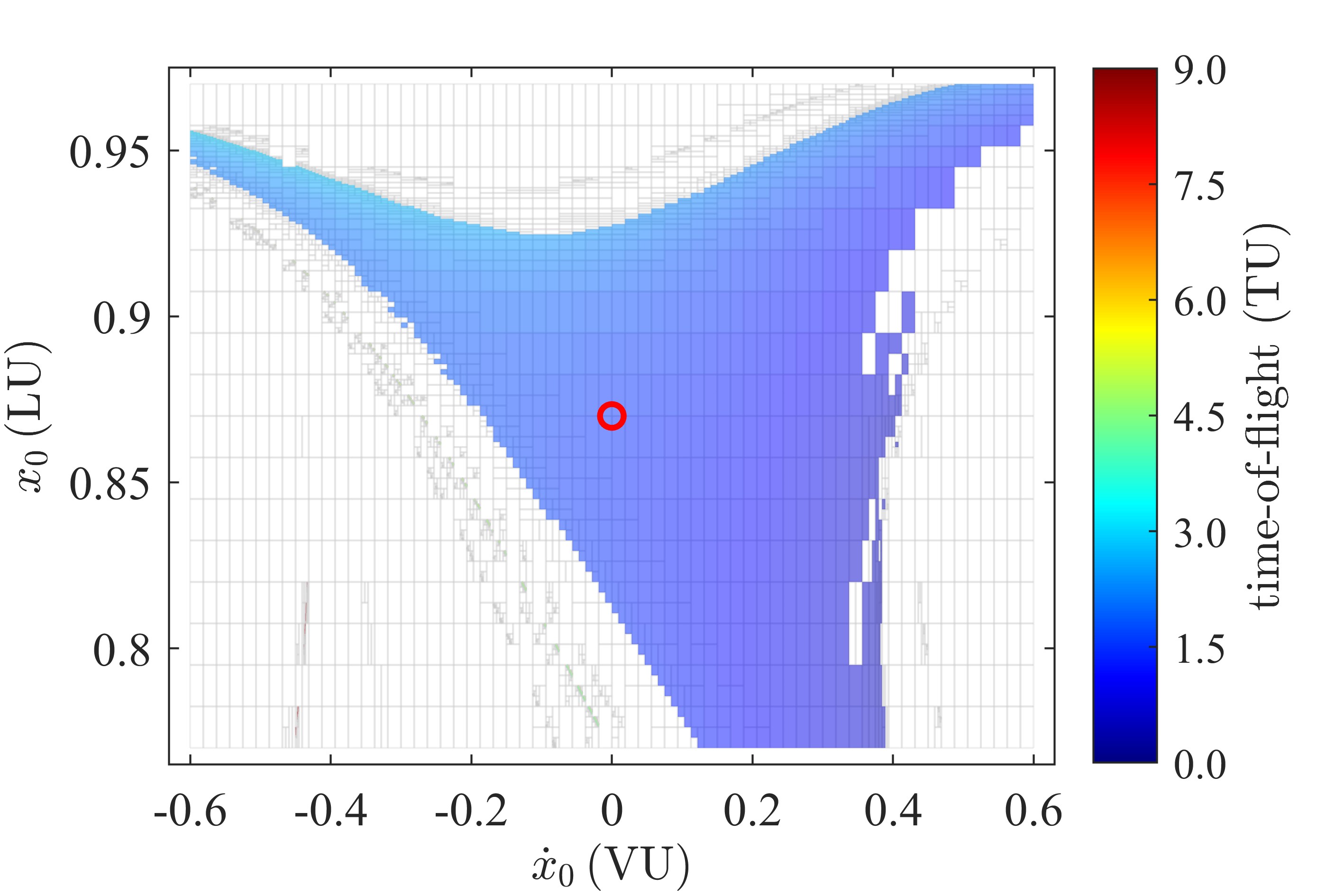}
    }
    \subfigure[Close up of \subref{fig4b}]{
        \label{fig4c}
        \includegraphics[width=0.48\linewidth]{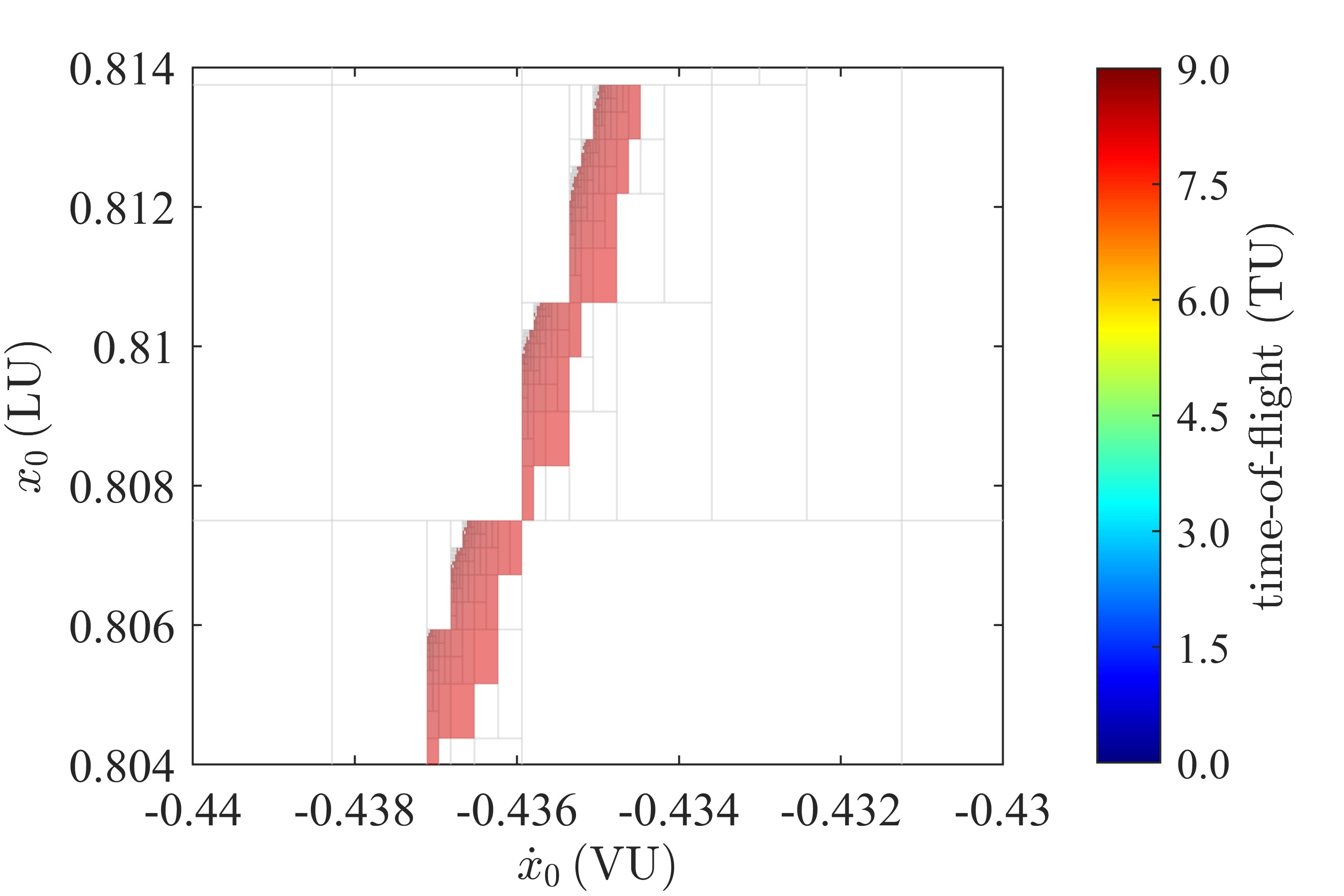}
    }
    \caption{\label{fig4} Subdomains obtained by the ADS process.}
\end{figure}

Using the ADS process in Algorithm~\ref{alg1}, 22,237 subdomains are obtained, of which 11,638 are feasible and 10,599 are infeasible (flagged by the LDB-based method), marked in blue and gray in Fig.~\ref{fig4a}, respectively.
Without incorporating the feasibility LDB-based detection strategy during the ADS process, we estimate that approximately 40,000 to 50,000 subdomains would have been generated, as an additional 20,000 to 30,000 infeasible regions would still undergo unnecessary splitting, resulting in more than twice the computational cost.

Figure~\ref{fig4b} shows the $\rm{ToF}$ associated with the central point of each feasible subdomain. It can be observed that most values are smaller than 4.5 TU, indicating that the majority of trajectories complete a single return to the Poincaré section within this time span. Only a small portion of the subdomains in the lower-left region in Fig.~\ref{fig4a} has a $\rm{ToF}$ greater than 6 TU, and it is displayed in Fig.~\ref{fig4c}. % These subdomains are extremely small in size, making them difficult to visually distinguish in Fig.~\ref{fig4b}.

\begin{figure}[!h]
    \centering
    \includegraphics[width=0.48\linewidth]{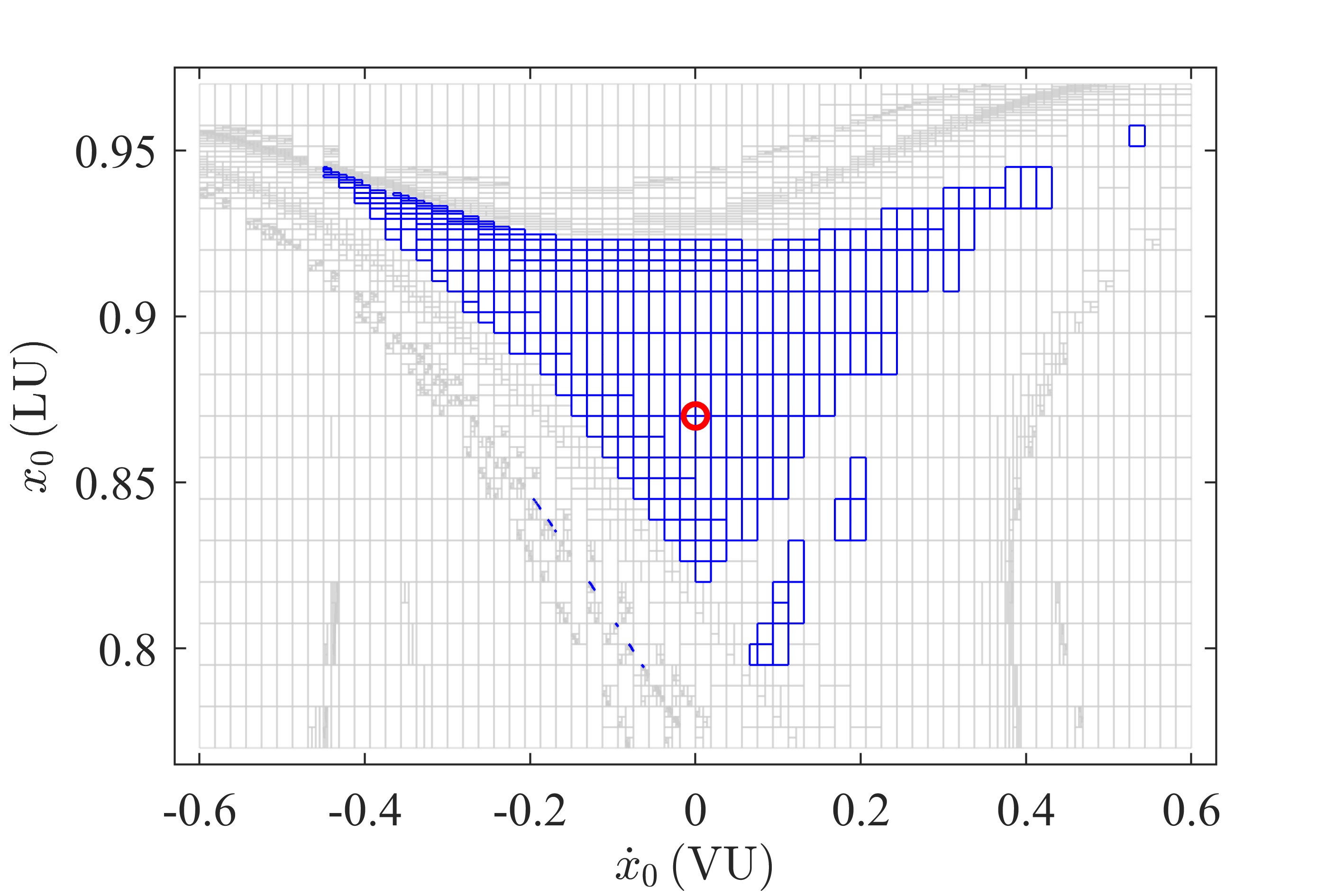}
    \caption{\label{fig5} Feasibility flagged by the first-stage POP.}
\end{figure}

At this point,  the approach described in Sec.~\ref{sec:Identification of Combinable HOTM Sequences} is employed to rigorously evaluate the combinability between all pairs of subdomains. This process eliminates subdomains that are incorrectly flagged as feasible by the LDB method and retains only those with verified feasibility. The refined set of feasible subdomains is shown in Fig.~\ref{fig5}. Moreover, the blue regions in Fig.~\ref{fig5} effectively represent the stable regions, where trajectories initiated within these subdomains return to the Poincaré section, indicating the presence of recurrent, stable motion. In contrast, the gray regions correspond to unstable regions, where trajectories fail to return, often due to escape, collision, or dynamical divergence. This refined partition thus provides meaningful insights into the underlying stability structure of the dynamical system.

\begin{figure}[!h]
    \centering
    \subfigure[DRO]{
        \label{fig6a}
        \includegraphics[width=0.48\linewidth]{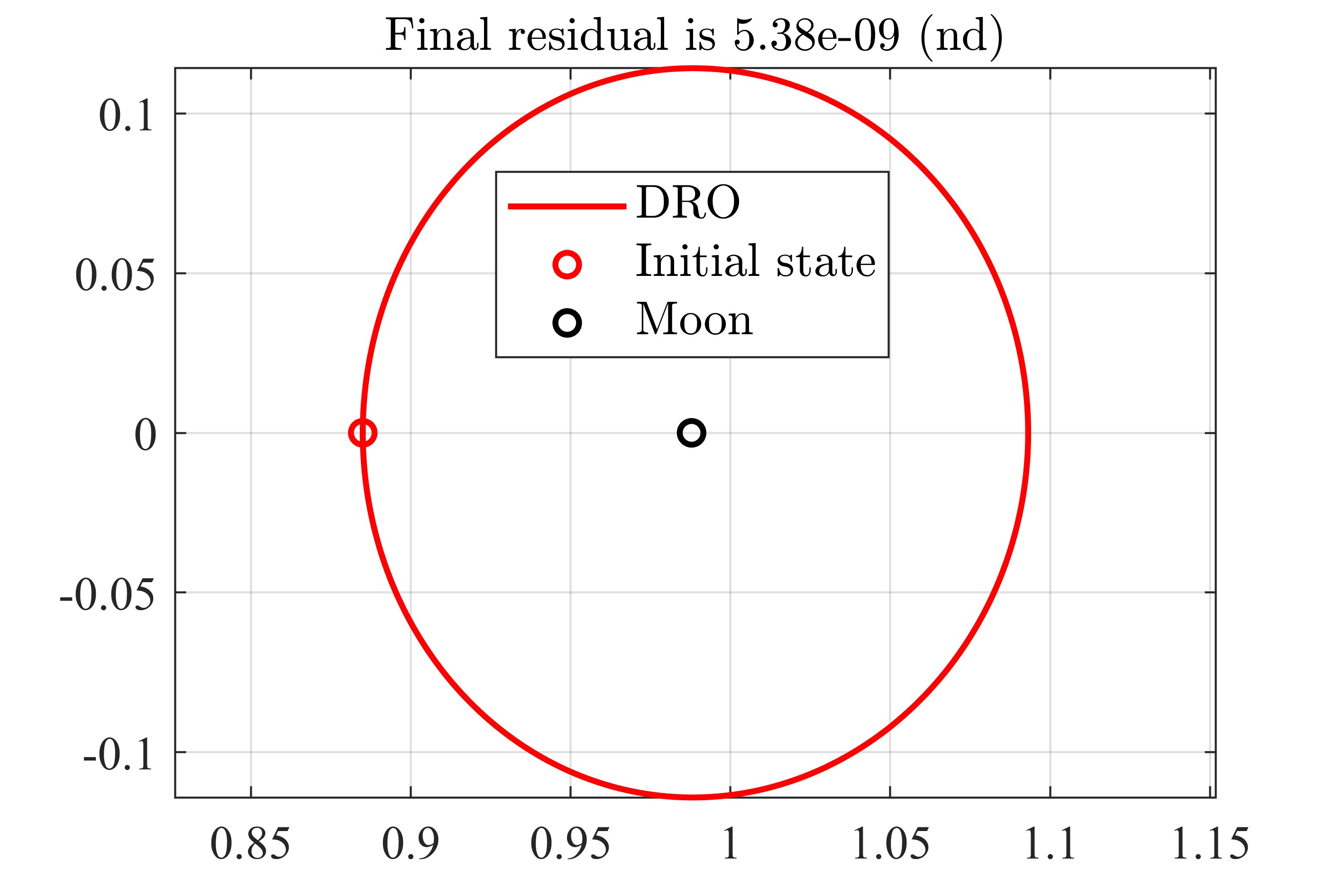}
    }
    \subfigure[P3DRO]{
        \label{fig6b}
        \includegraphics[width=0.48\linewidth]{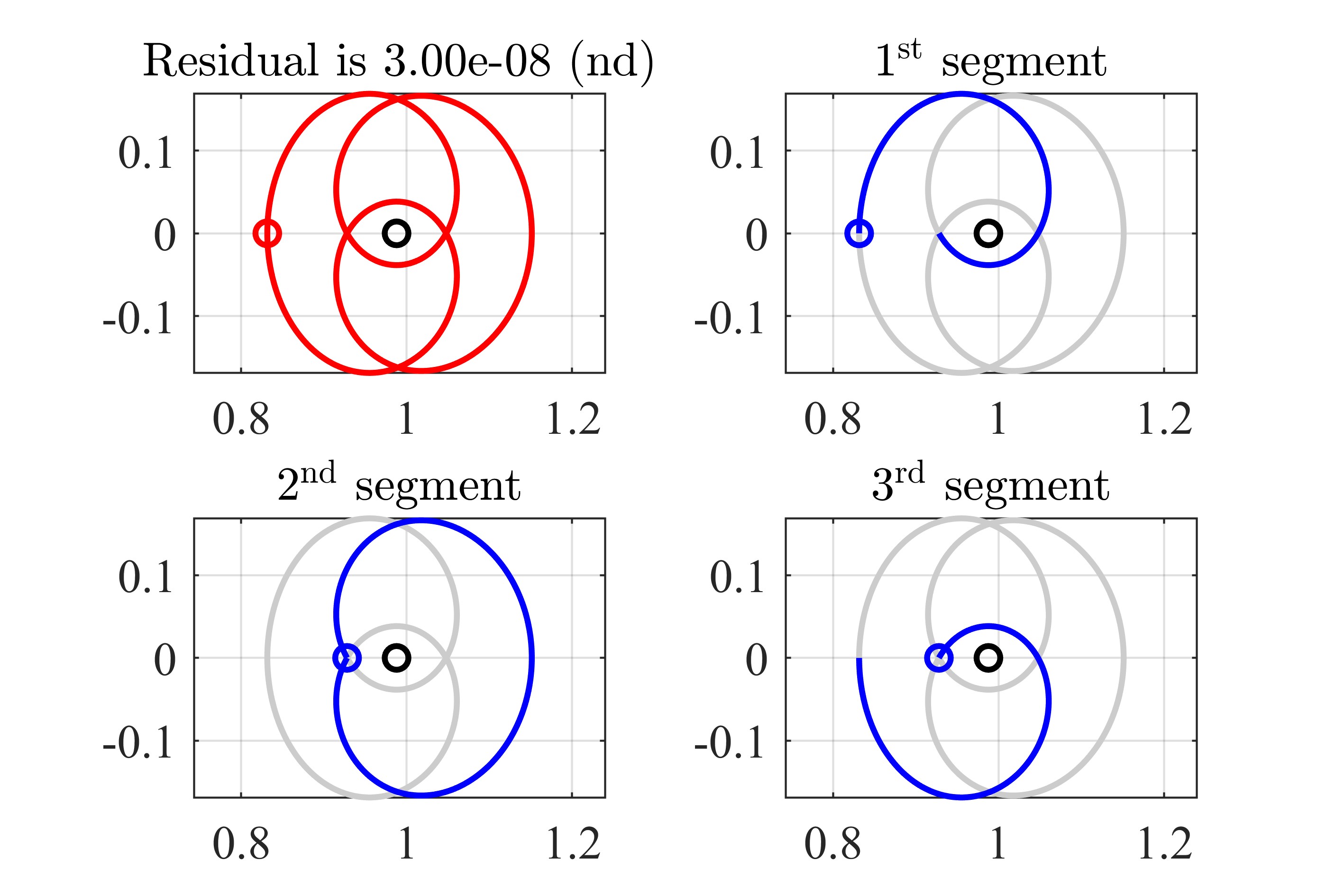}
    }
    \caption{\label{fig6} DRO and P3DRO in the first planar case.}
\end{figure}

\begin{figure}[!h]
    \centering
    \subfigure[Type I]{
        \label{fig8a}
        \includegraphics[width=0.48\linewidth]{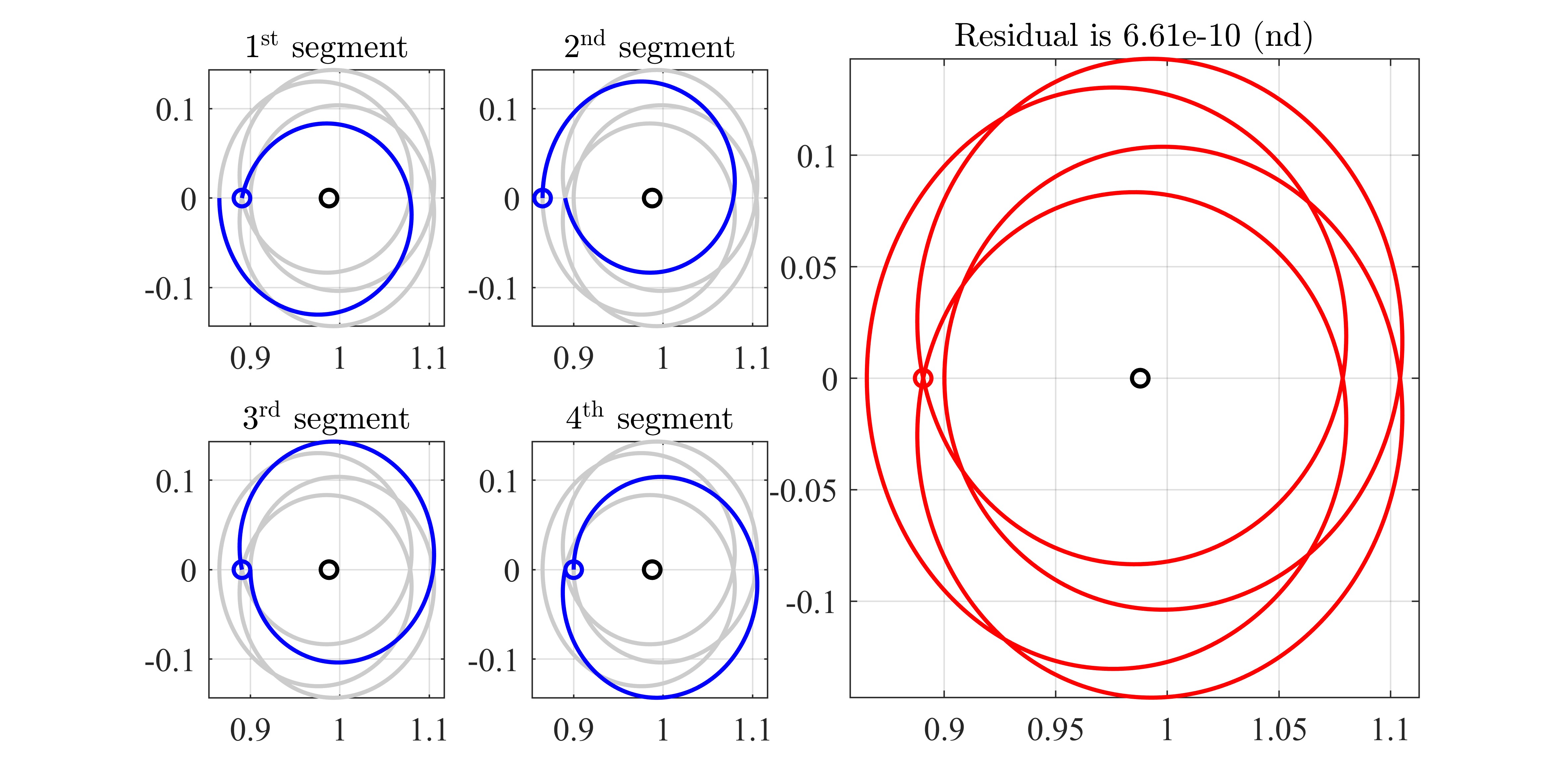}
    }
    \subfigure[Type II]{
        \label{fig8b}
        \includegraphics[width=0.48\linewidth]{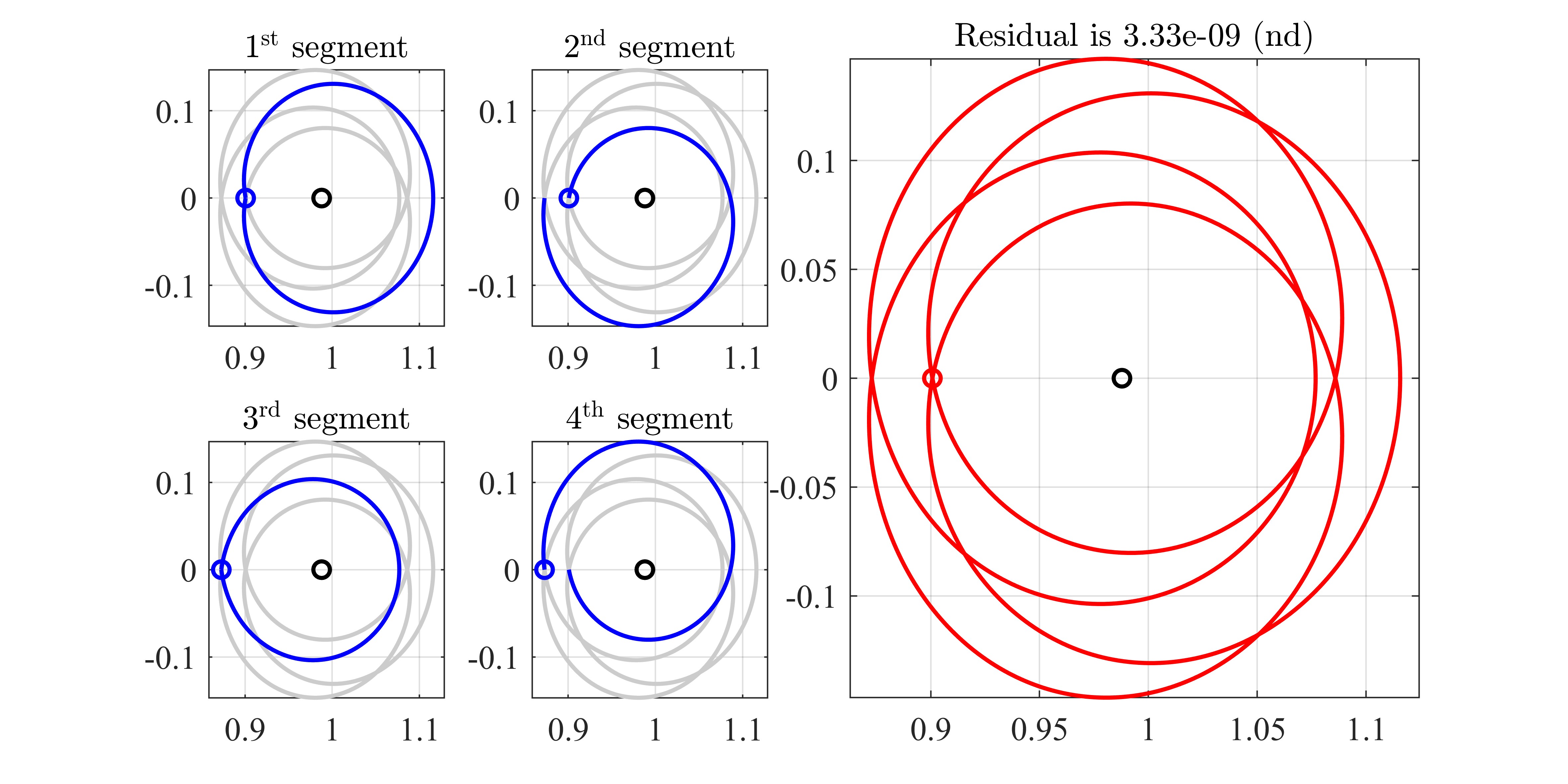}
    }
    \caption{\label{fig8} P4DROs in the first planar case.}
\end{figure}

\begin{figure}[!h]
    \centering
    \includegraphics[width=0.5\linewidth]{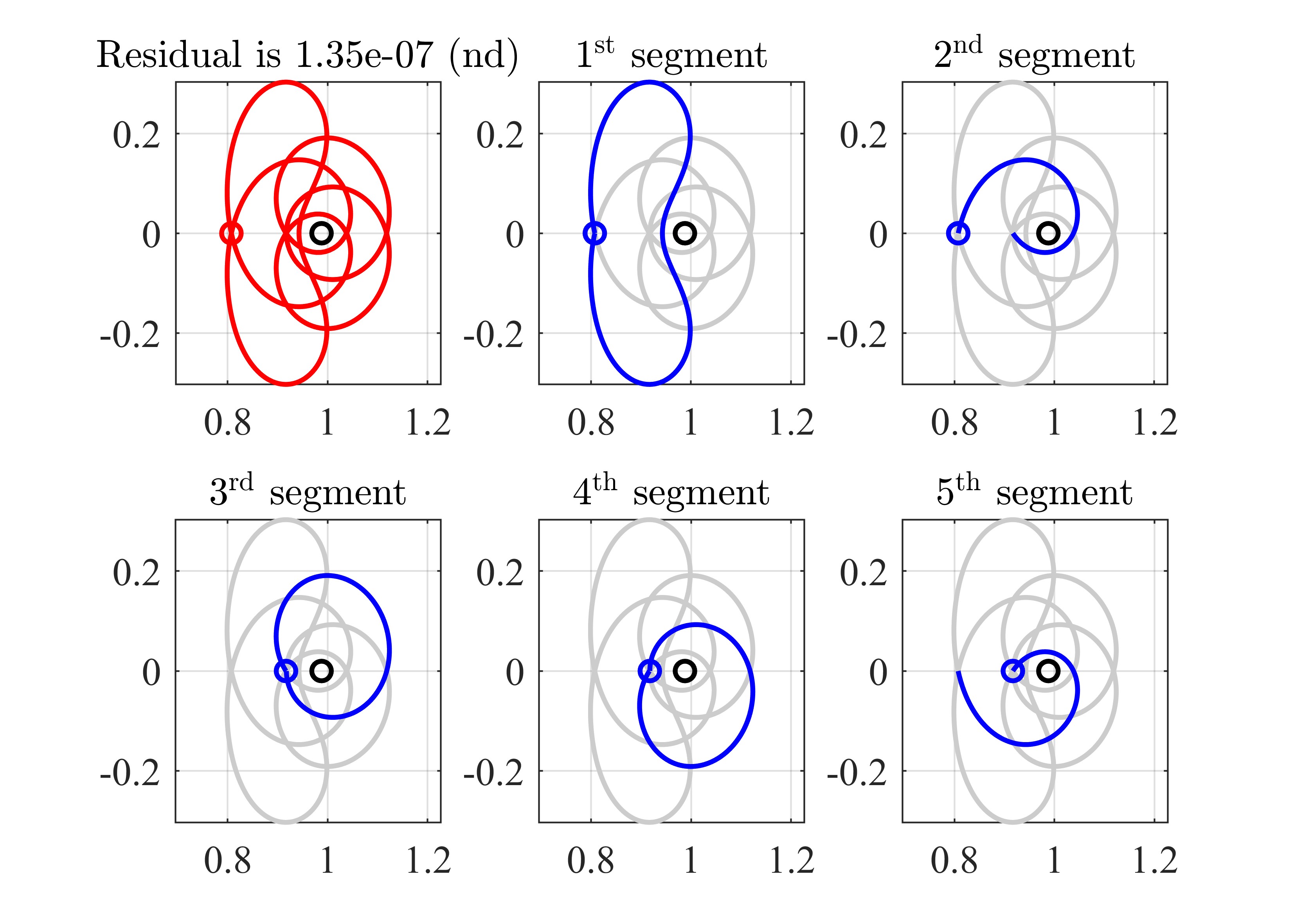}
    \caption{\label{fig9} P5g' in the first planar case.}
\end{figure}

\begin{figure}[!h]
    \centering
    \subfigure[Type I]{
        \label{fig10a}
        \includegraphics[width=0.48\linewidth]{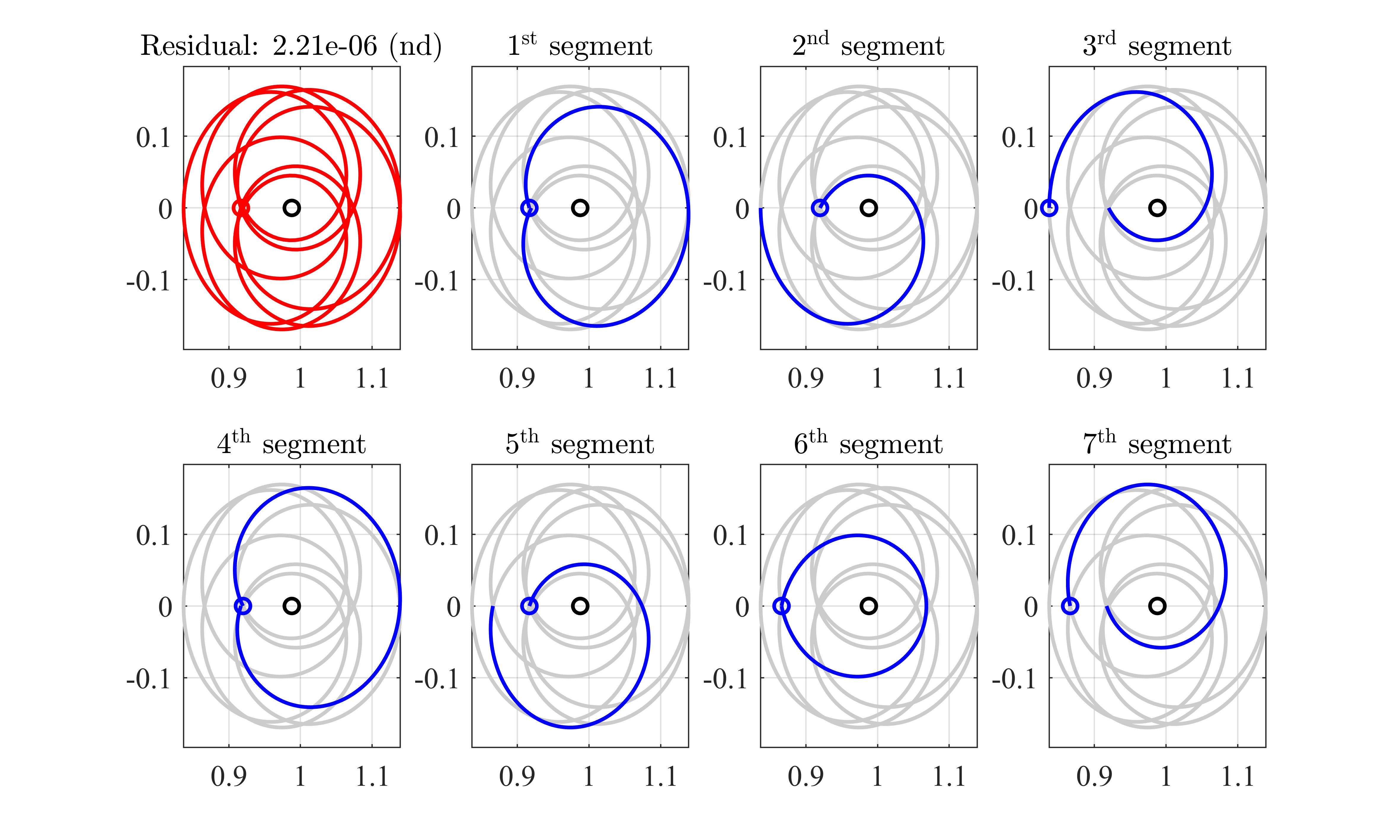}
    }
    \subfigure[Type II]{
        \label{fig10b}
        \includegraphics[width=0.48\linewidth]{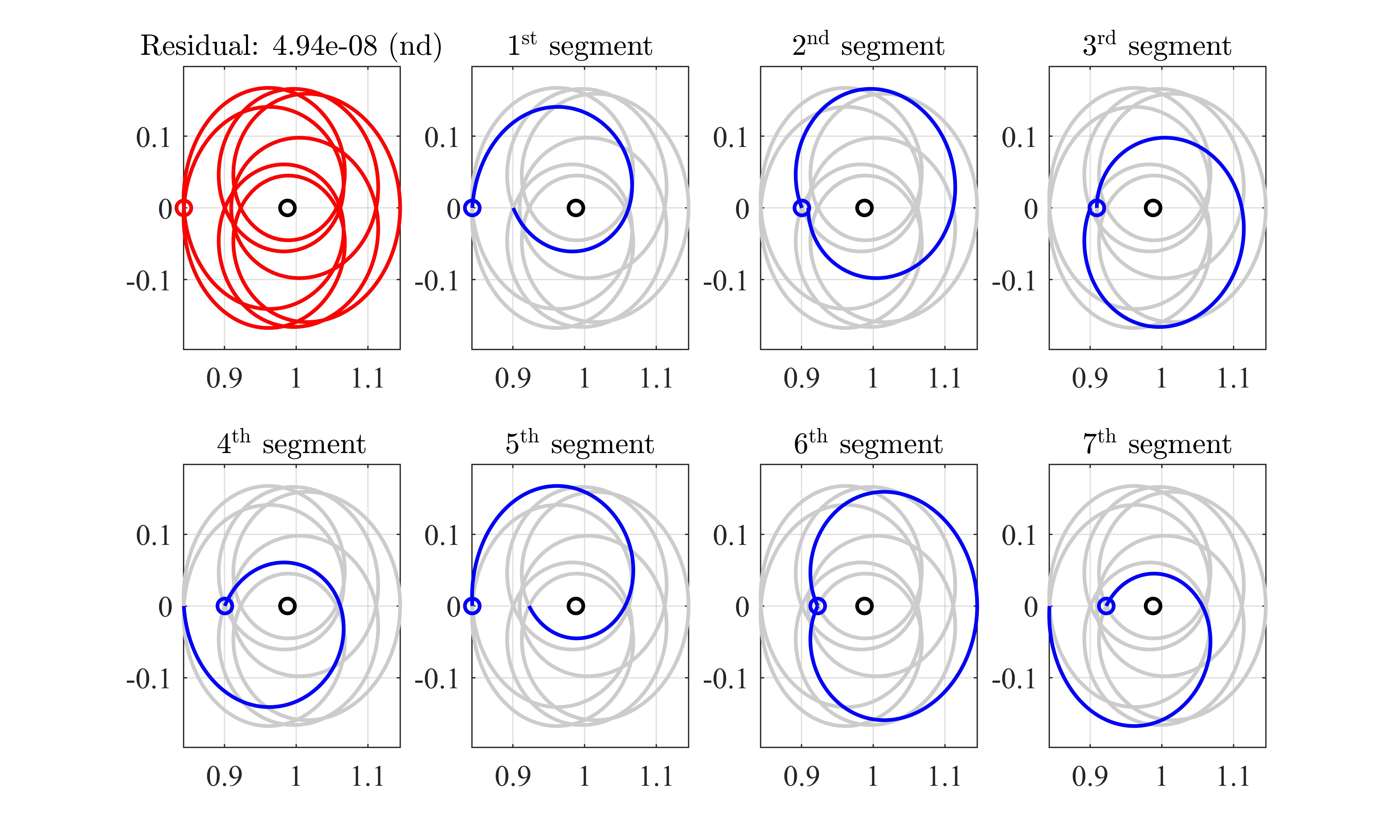}
    }
    \caption{\label{fig10} P7DROs in the first planar case.}
\end{figure}

\begin{figure}[!h]
    \centering
    \subfigure[P7g']{
        \label{fig11a}
        \includegraphics[width=0.48\linewidth]{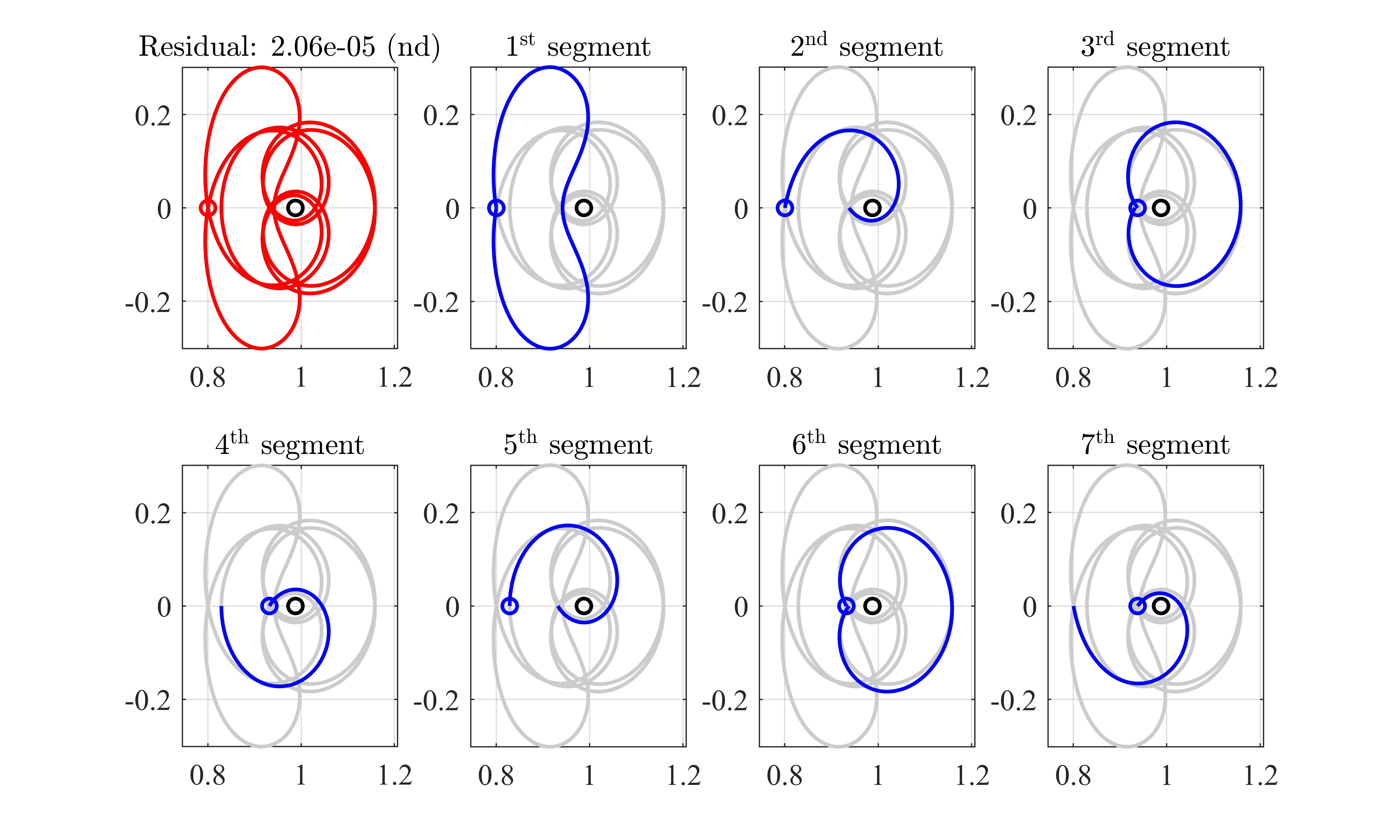}
    }
    \subfigure[P9g']{
        \label{fig11b}
        \includegraphics[width=0.48\linewidth]{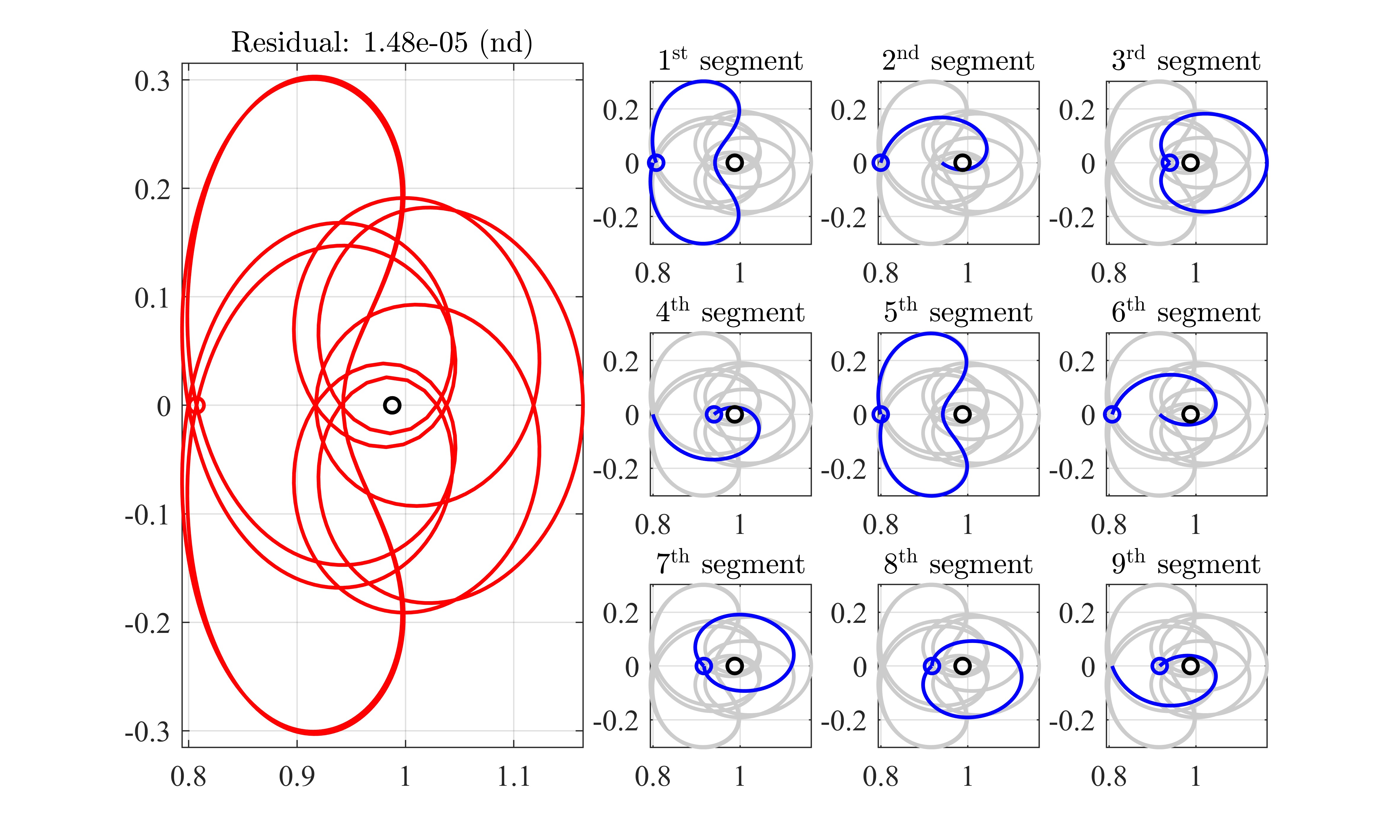}
    }
    \caption{\label{fig11} P7g' and P9g' in the first planar case.}
\end{figure}

Table~\ref{tab6} presents the fixed points identified in the planar CRTBP case, covering POs ranging from 1 to 9 revolutions. The corresponding periodic trajectories are plotted in Figs.~\ref{fig6a}-\ref{fig11b}, as indicated in the last column of Table~\ref{tab6}. For multi-revolution POs in Figs.~\ref{fig6b}-\ref{fig11b}, both the complete trajectory (red curve) and each individual revolution segment (blue curves in subplots) are illustrated. The initial state of each segment is marked with a circular marker for clarity. Several key points should be noted:
\begin{enumerate}
    \item All listed fixed points are obtained through RPO. The trajectories in the corresponding figures are generated by directly propagating the polynomial solutions. Due to the inherent approximation errors in the HOTMs, small mismatches may occur between the start and end points of the trajectories. These residuals are marked in each figure. % (see Figs.~\ref{fig6a}-\ref{fig11b}).
    In principle, these polynomial-based trajectories can be refined to yield closed POs using standard differential correction techniques.
    \item For the one-revolution case, while Lyapunov orbits also qualify as single-revolution periodic solutions, no such orbit is identified under the given Jacobi constant $C_J=3.00022$ because the initial condition of the Lyapunov orbit ($x_0=0.7688974950452078$, $\dot{x}_0=0$, and $\dot{y}_0=0.4811655138591737$) lies outside the specified initial domain in Eq.~\eqref{eq50}.
    \item For higher-revolution results, repeated traversals of a single-revolution DRO (\emph{e.g.}, $2\times$DRO, $3\times$DRO) also mathematically qualify as multi-revolution solutions. However, to focus on nontrivial solutions, such duplicated DRO-based orbits are excluded from the final list (see Table~\ref{tab6}).
    \item As the number of revolutions increases, the residual between the start and end states in the polynomial solution tends to grow, reflecting the accumulation of approximation error with increasing trajectory length.
    \item For DRO and P3DRO solutions in Figs.~\ref{fig6a} and \ref{fig8}), respectively, the theoretical $\dot{x}_0$ should be exactly zero. However, due to numerical issues in the RPO optimization process, the obtained value is a small, non-zero number on the order of $10^{-12}$. In Table~\ref{tab6}, we adopt a convention where any value with an absolute magnitude smaller than $10^{-10}$ is reported as zero.
\end{enumerate}

\begin{table}[!h]
    \caption{\label{tab6} Valid fixed points obtained in the first planar case}
    \centering
    \begin{tabular}{lllllll}
    \hline\hline
    \multirow{2}{*}{$n$} & \multirow{2}{*}{PO Type} & \multicolumn{3}{l}{Initial state} & \multirow{2}{*}{Period (TU)} & \multirow{2}{*}{Figure} \\ \cline{3-5}
    & & $x_0$ (LU) & $\dot{x}_0$ (VU) & $\dot{y}_0$ (VU) & & \\ \hline
    1 & DRO & 0.885009684799908 & 0 & 0.470630255988515 & 1.57454364775953 & \ref{fig6a} \\
    3 & P3DRO & 0.831591486122089 & 0 & 0.434090978642371 & 5.23081766916226 & \ref{fig6b} \\
    4 & P4DRO-I & 0.89052832582472 & 0.0985878312870881  & 0.470646557373107 & 6.37463624124285 & \ref{fig8a} \\
    & P4DRO-II & 0.900772797217838 & -0.0927028710649920 & 0.495783314283405 & 6.39078214289561 & \ref{fig8b} \\
    5 & P5g’ & 0.807357887647950 & -0.0956081545978604 & 0.433518861583397 & 11.1751086919436 & \ref{fig9} \\
    7 & P7DRO-I & 0.916929181700578 & -0.175717632213615 & 0.528042521610021 & 11.6960585356669 & \ref{fig10a} \\
    & P7DRO-II & 0.843301779064331 & 0.0264518382948230 & 0.433467279267690 & 11.6716427018773 & \ref{fig10b} \\
    & P7g' & 0.800533327625822 & -0.0783668113509505 & 0.441933007711387 & 15.0692523706070 & \ref{fig11a} \\
    9 & P9g’ & 0.807337935300132 & -0.0956506138795539 & 0.433522872600990 & 20.9914771396290 & \ref{fig11b}           
    \\ \hline\hline
    \end{tabular}
\end{table}

The obtained P\emph{n}DROs (\emph{e.g.}, P3DRO from Fig.~\ref{fig6b}, P4DROs from Fig.~\ref{fig8}, and P7DROs from Fig.~\ref{fig10}) are bifurcating families of the DRO family. In addition, a family of previously undocumented POs has been revealed and has been here denoted as P\emph{n}g'.
These POs are characterized by their hybrid structure, typically consisting of at least one segment that resembles a Lyapunov-type trajectory and another resembling a DRO-type segment. Such a configuration suggests that P\emph{n}g' orbits may arise from high-periodicity bifurcations related to the g' family, as described in Hénon’s dynamical classification \cite{Henon2003}. The structural composition of these orbits suggests their potential as dynamical bridges between DRO and Lyapunov orbits, offering new possibilities for mission design that involve transitions between different orbit types. This further demonstrates the proposed method’s capability to uncover novel orbital behaviors without prior knowledge or continuation procedures.

All fixed points presented in Table~\ref{tab6} are further refined by differential correction, specifically by adjusting the half-period velocity component $\dot{y}$ to enforce a near-zero value. This procedure ensures enhanced symmetry and improves the periodicity of the trajectory. The differential correction converged successfully for all cases. After correction, the half-period $\dot{y}$ is reduced to approximately $10^{-14}$, approaching machine precision. When the corrected state propagates for a complete period, the mismatch between the initial and terminal states is generally of the order of $10^{-12}$. For the P9g' orbit, which involves a longer propagation time (20.99 TU, which is approximately three months in the Earth-Moon system) and close lunar encounters, the final residual is approximately $10^{-10}$.

\begin{figure}[!h]
    \centering
    \subfigure[P2g'-like]{
        \label{fig13a}
        \includegraphics[width=0.48\linewidth]{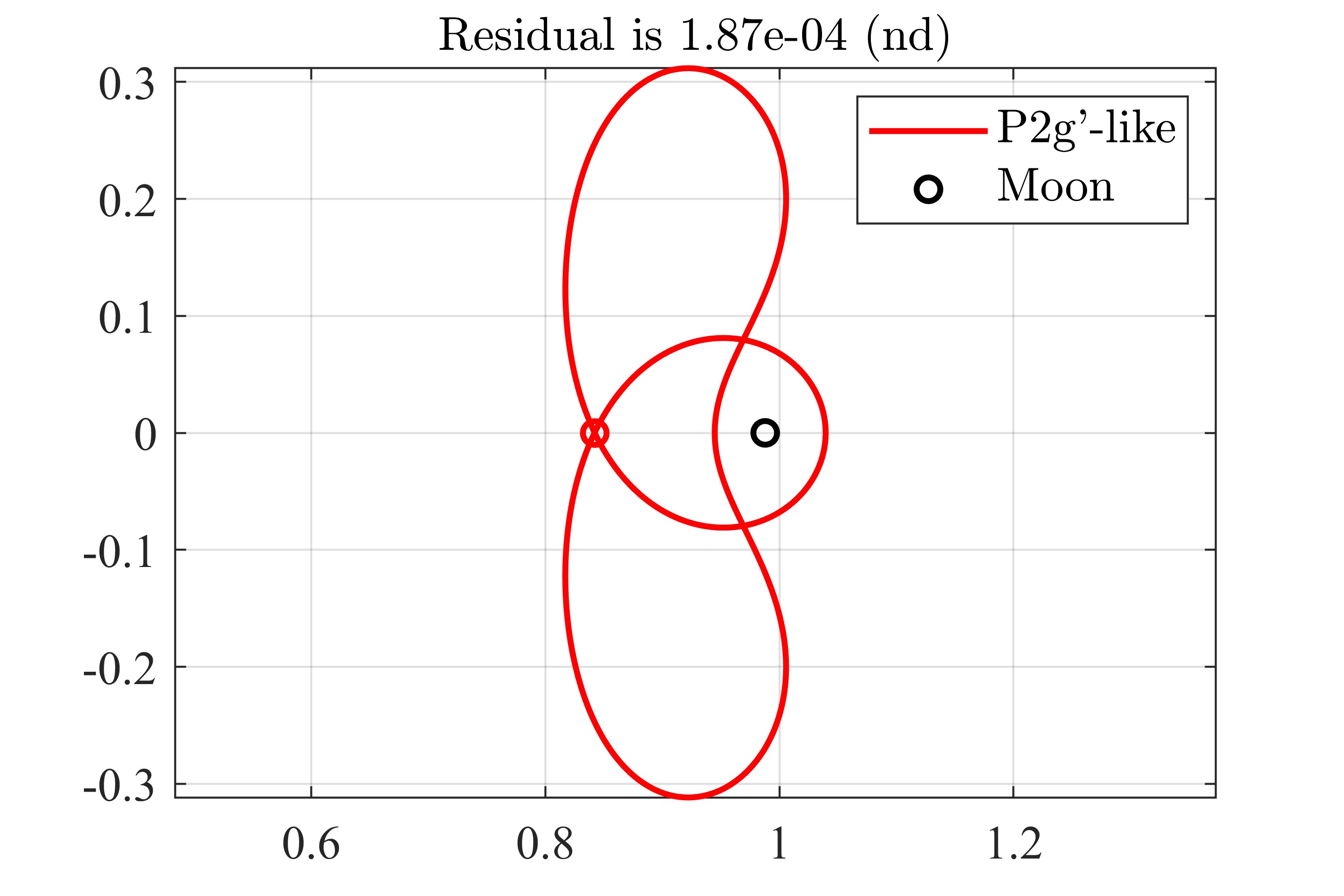}
    }
    \subfigure[P4g'-like]{
        \label{fig13b}
        \includegraphics[width=0.48\linewidth]{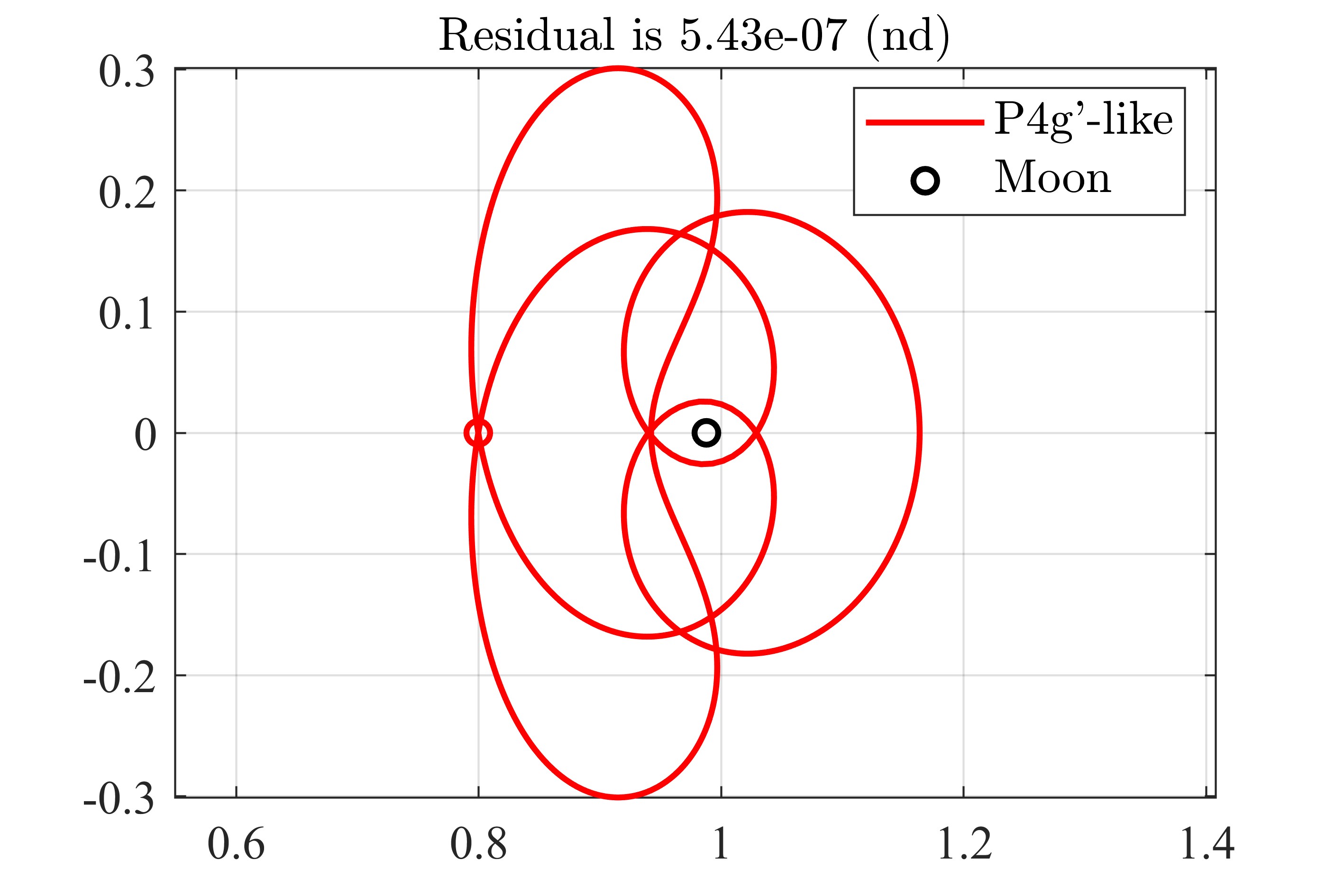}
    }
    \caption{\label{fig13} P2g'- and P4g'-like orbits in the first planar case.}
\end{figure}

In addition to the fixed points listed in Table~\ref{tab6}, several seemingly favorable solutions are also identified, characterized by relatively small residuals after RPO-based optimization. However, these solutions do not qualify as true fixed points, as they either fail to converge under differential correction or eventually collapse onto the known DRO solution % (\emph{i.e.}, Fig.~\ref{fig6a})
during the correction process. Two representative examples are summarized in Table~\ref{tab7}, and represented in Figs.~\ref{fig13a}-\ref{fig13b}. The term "P\emph{n}g'-like" is used to describe these orbits because they exhibit structural features similar to the P\emph{n}g' family listed in Table~\ref{tab6}. Although these trajectories are not strictly periodic, they can be converted into true POs through the application of small corrective maneuvers. Based on the above results, it is speculated that the newly discovered P\emph{n}g' family may exist only for odd values of \emph{n}.

\begin{table}[!h]
    \caption{\label{tab7} Invalid fixed points obtained in the first planar case}
    \centering
    \begin{tabular}{lllllll}
    \hline\hline
    \multirow{2}{*}{$n$} & \multirow{2}{*}{PO Type} & \multicolumn{3}{l}{Initial state} & \multirow{2}{*}{ToF (TU)} & \multirow{2}{*}{Figure} \\ \cline{3-5}
    & & $x_0$ (LU) & $\dot{x}_0$ (VU) & $\dot{y}_0$ (VU) & & \\ \hline
    2 & P2g'-like & 0.842265625000290 & -0.187664348385473 & 0.391450394121855 & 5.95981828071091 & \ref{fig13a} \\
    4 & P4g'-like & 0.799721494787247 & 0.0763263625927055 & 0.442936120657196 & 9.81632980112884 & \ref{fig13b} \\
    \hline\hline
    \end{tabular}
\end{table}

\subsubsection{Second Planar Case} \label{sec:Second Planar Case}
\begin{figure}[!h]
    \centering
    \subfigure[DRO and Lyapunov orbit]{
        \label{fig15a}
        \includegraphics[width=0.48\linewidth]{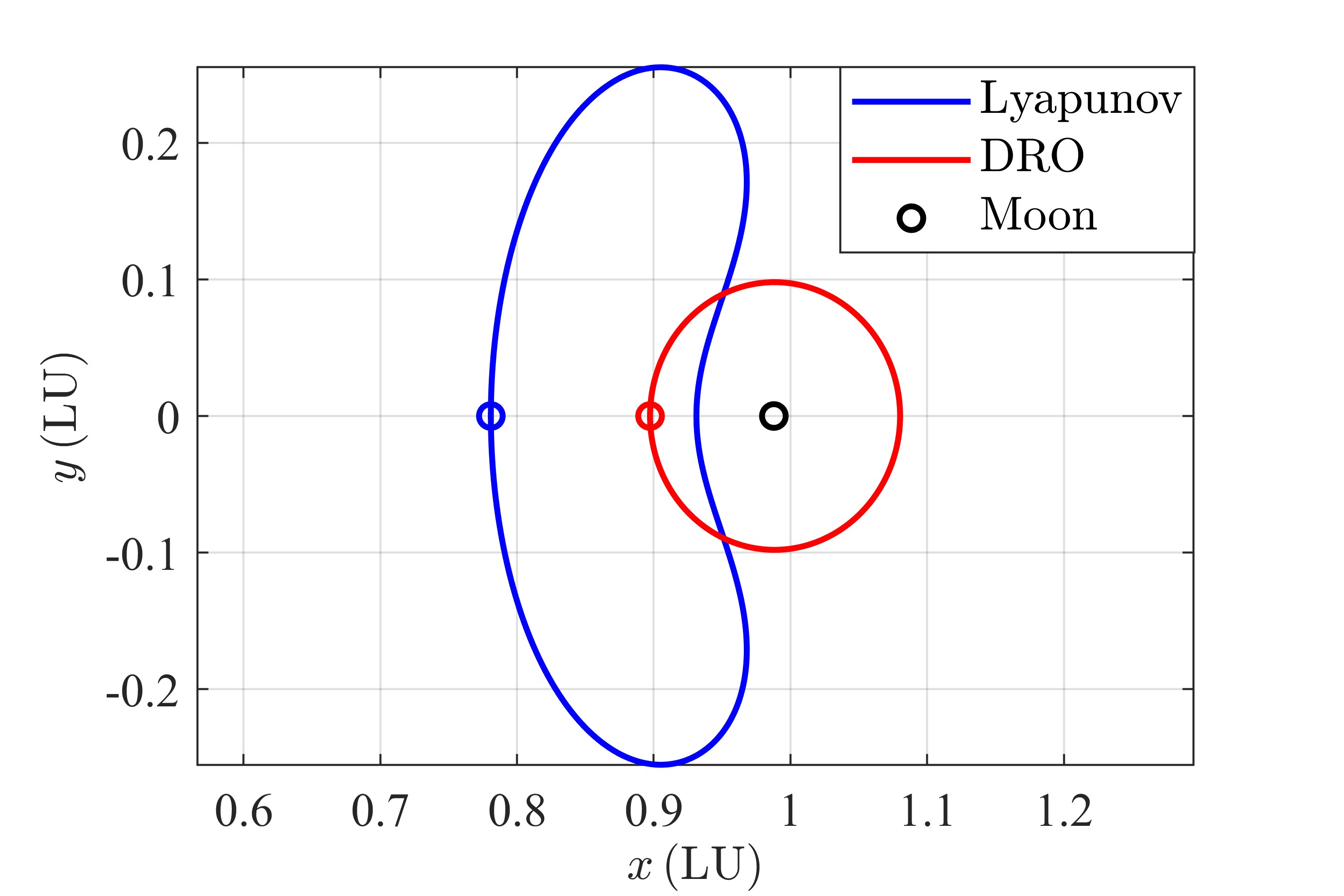}
    }
    \subfigure[P3g']{
        \label{fig15b}
        \includegraphics[width=0.48\linewidth]{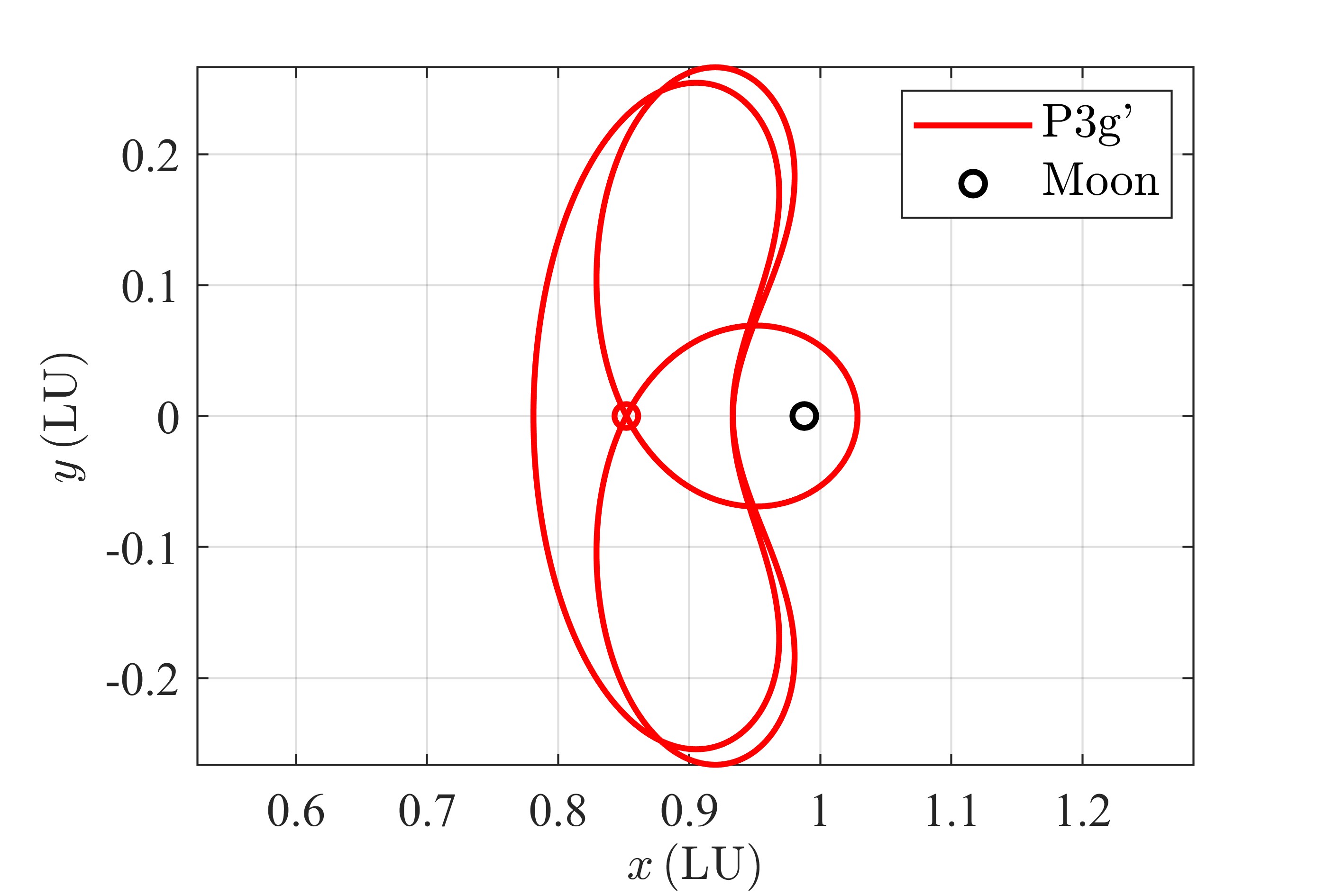}
    }
    \caption{\label{fig15} Valid POs obtained in the second planar case.}
\end{figure}

A second planar case is investigated herein. This case has an initial domain of 
\begin{equation} \label{eq51}
    {{\mathscr D}_0} = [0.56,0.96] \times [ - 0.2,0.2] \,,
\end{equation}
with a Jacobi constant of 3.020052, and a maximal time-of-flight of 25. Both DRO and Lyapunov orbits are found, as provided in Table~\ref{tab8} and Fig.~\ref{fig15a}. 
In addition, a P3g' orbit, missed in the first planar case due to the limited search domain, is also found here, as shown in Fig.~\ref{fig15b}.

\begin{table}[!h]
    \caption{\label{tab8} Valid fixed points obtained in the second planar case}
    \centering
    \begin{tabular}{lllllll}
    \hline\hline
    \multirow{2}{*}{$n$} & \multirow{2}{*}{PO Type} & \multicolumn{3}{l}{Initial state} & \multirow{2}{*}{Period (TU)} & \multirow{2}{*}{Figure} \\ \cline{3-5}
    & & $x_0$ (LU) & $\dot{x}_0$ (VU) & $\dot{y}_0$ (VU) & & \\ \hline
    1 & DRO & 0.897304549432615 & 0 & 0.475278703786673 & 1.32961331350808 & \ref{fig15a} \\
    & Lyapunov & 0.780896659202750 & 0 & 0.445473137340473 & 3.97137299745678 & \ref{fig15a} \\
    3 & P3g' & 0.852098052983502 & -0.187721536949396 & 0.368541113320107 & 9.25206400352203 & \ref{fig15b} \\   
    \hline\hline
    \end{tabular}
\end{table}

\begin{figure}[!h]
    \centering
    \subfigure[$n=2$]{
        \label{fig16a}
        \includegraphics[width=0.48\linewidth]{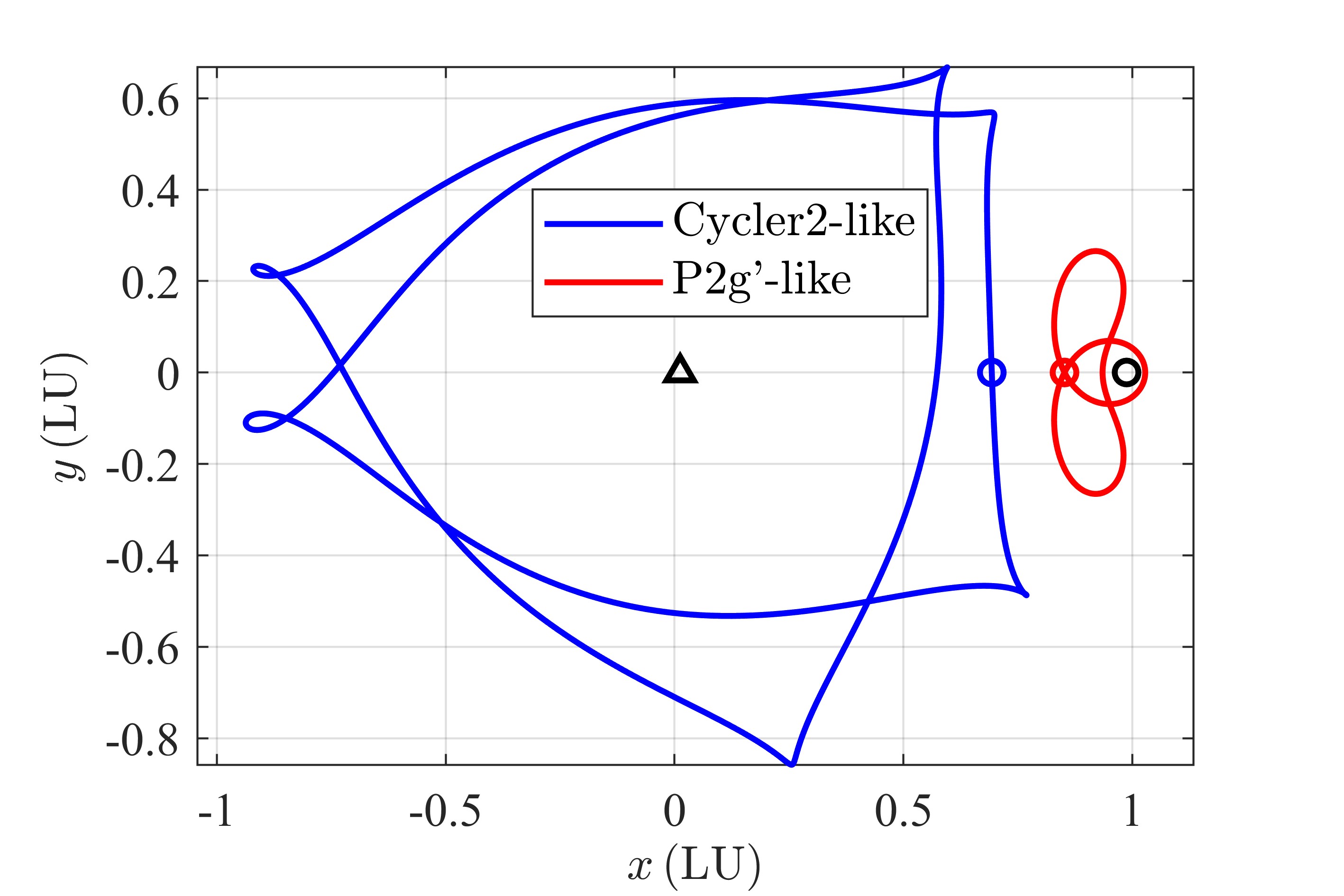}
    }
    \subfigure[$n=3$]{
        \label{fig16b}
        \includegraphics[width=0.48\linewidth]{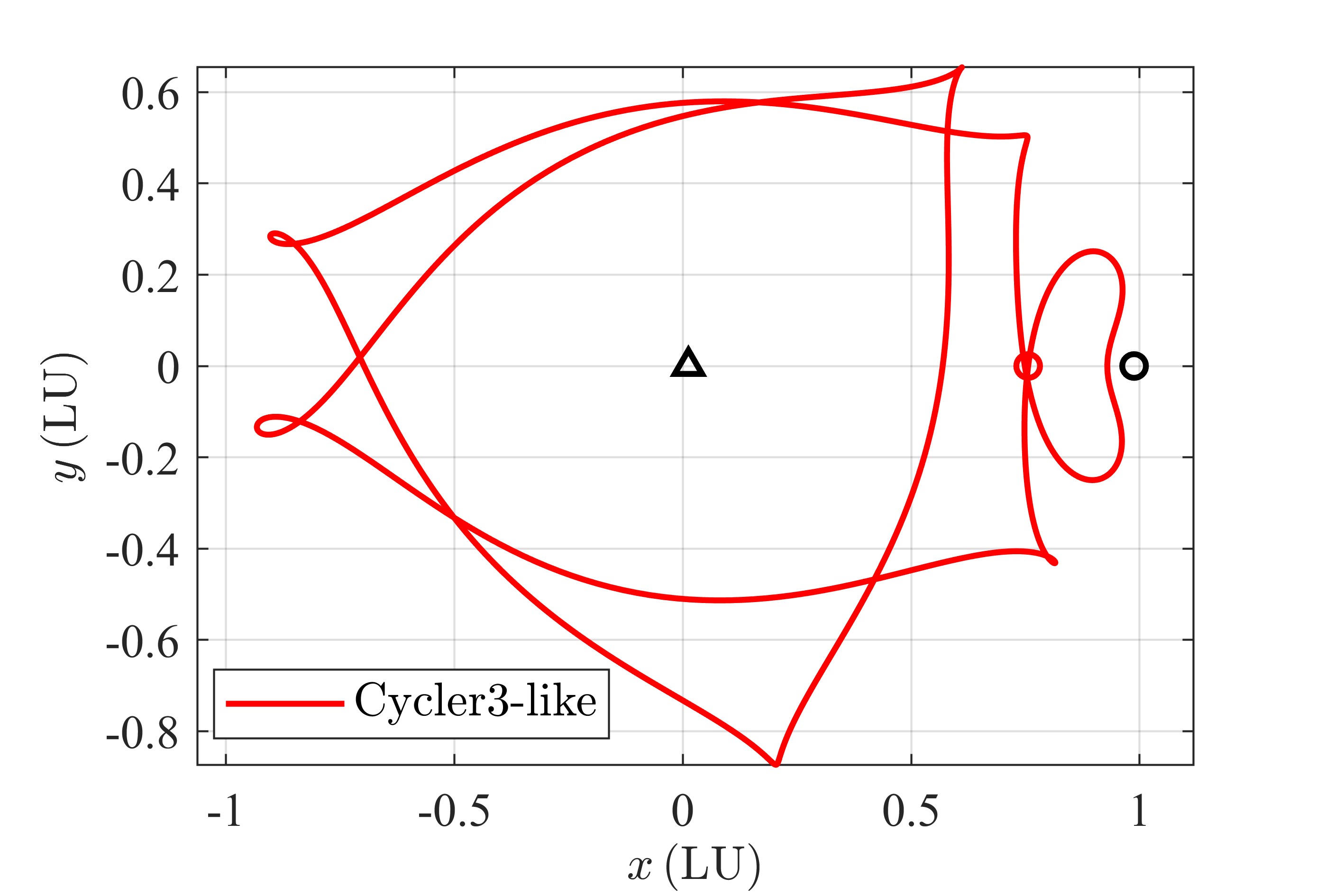}
    }
    \subfigure[$n=4$]{
        \label{fig16c}
        \includegraphics[width=0.48\linewidth]{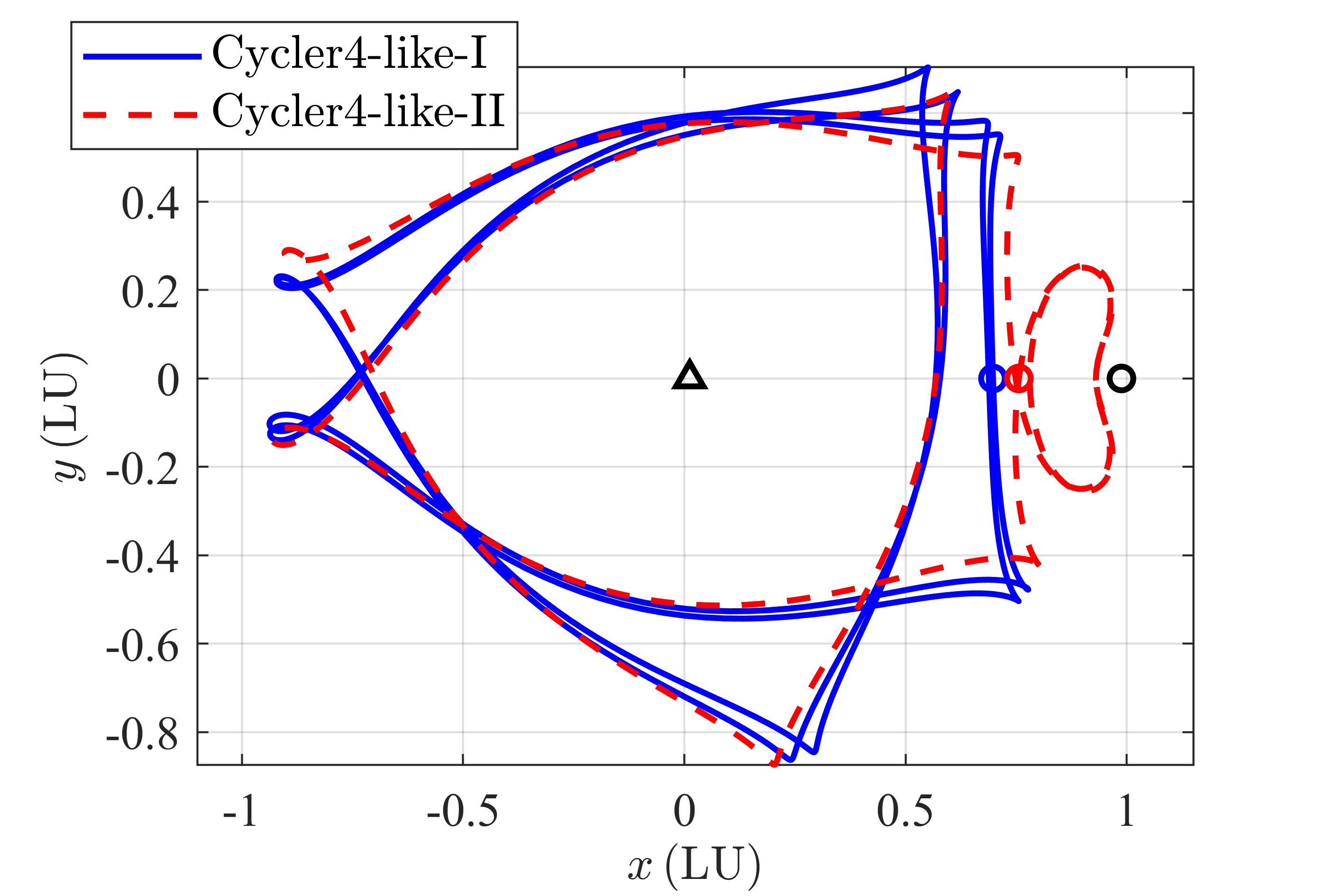}
    }
    \subfigure[$n=5$]{
        \label{fig16d}
        \includegraphics[width=0.48\linewidth]{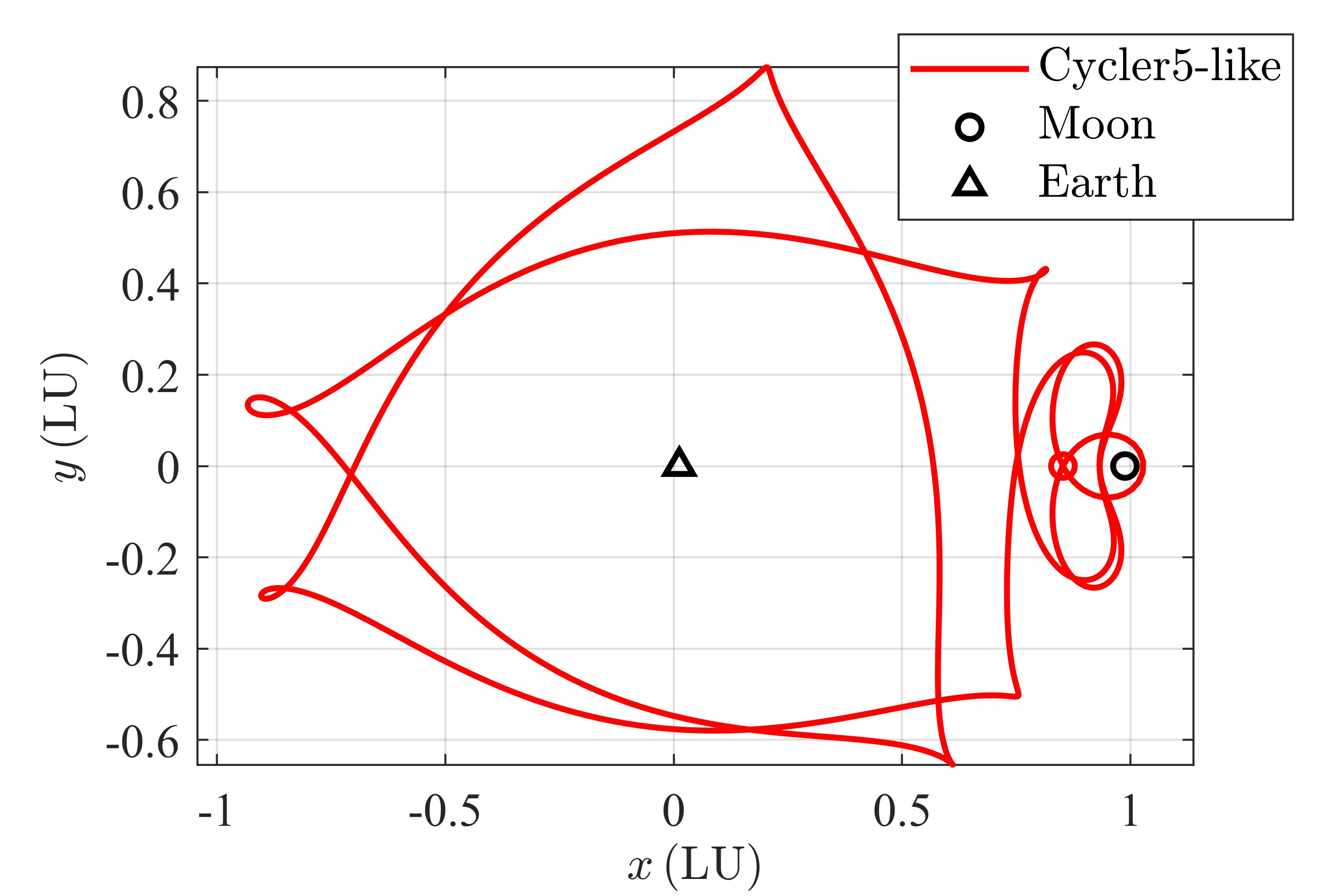}
    }
    \caption{\label{fig16} Invalid POs obtained in the second planar case.}
\end{figure}

Additionally, a number of invalid solutions are identified during the second planar case and detailed in Table~\ref{tab9}. Among them is a P2g'-like orbit, as shown in Fig.~\ref{fig16a} (see also Fig.~\ref{fig13a} for the first planar case). Moreover, by expanding the initial search domain (particularly the range of $x_0$) and increasing the allowable maximum time-of-flight, several new POs are found and are displayed in Fig.~\ref{fig16}. These trajectories shuttle between both sides of the Earth, with apogees being approximately around -1, and maintain relatively large distances from both the Earth and the Moon, which contributes to their enhanced stability. 
These orbits resemble the Earth–Moon cycler trajectories reported in the literature \cite{Ross2025cycler}, although they are not strictly periodic, as they cannot be refined through differential correction. Nevertheless, similar to the previous invalid solutions in Table~\ref{tab7}, they can be converted into true POs by applying extremely small control maneuvers.

\begin{table}[!h]
    \caption{\label{tab9} Invalid fixed points obtained in the second planar case}
    \centering
    \begin{tabular}{lllllll}
    \hline\hline
    \multirow{2}{*}{$n$} & \multirow{2}{*}{PO Type} & \multicolumn{3}{l}{Initial state} & \multirow{2}{*}{ToF (TU)} & \multirow{2}{*}{Figure} \\ \cline{3-5}
    & & $x_0$ (LU) & $\dot{x}_0$ (VU) & $\dot{y}_0$ (VU) & & \\ \hline
    2 & Cycler2-like & 0.693257903603195 & -0.0211935974664688 & 0.585989070384689 & 23.5015611948510 & \ref{fig16a} \\
    & P2g'-like & 0.852212540121389 & -0.187499999999992 & 0.368712771276019 & 5.27925556899687 & \ref{fig16a} \\
    3 & Cycler3-like & 0.755917269316675 & 0.0607604069115931 & 0.474068882807608 & 27.3881708212492 & \ref{fig16b} \\
    4 & Cycler4-like-I & 0.697519898839708 & -0.0192404589366478 & 0.577737146771673 & 47.0084870811382 & \ref{fig16c} \\
    & Cycler4-like-II & 0.756036541995036 & 0.0609453131165504 & 0.473870929936850 & 31.35979268454441 & \ref{fig16c} \\ 
    5 & Cycler5-like & 0.852074288972938 & -0.187890624999752 & 0.368442765173470 & 32.6703329457782 & \ref{fig16d} \\
    \hline\hline
    \end{tabular}
\end{table}

%%%%%%%%%%%%%%%%%%%%%%%%%%%%%%%%%%%%%%%%%%%%%%%%%%%%%%%%%%%%%%%%%%%%%%%%%%%%%%%%%%%%%%%%%%%
\subsection{Spatial Case Results} \label{sec:Spatial Case Results}
The central point of the spatial case is selected as ${x_0} = 0.84$, ${\dot x_0} = 0$, ${z_0} = 0.05$, and ${\dot z_0} = 0$, with the Jacobi constant being $C_J=2.9519$. The half-lengths of the initial domain ${{\mathscr D}_0}$ are set as $\Delta {x_0} = 0.04$, $\Delta {\dot x_0} = 0.08$, $\Delta {z_0} = 0.05$,  and $\Delta {\dot z_0} = 0.16$; thus, the initial domain ${{\mathscr D}_0}$ of the spatial case in Eq.~\eqref{eq27} is expressed as
\begin{equation} \label{eq52}
    {{\mathscr D}_0} = [0.8,0.88] \times [ - 0.08,0.08] \times [0.0,0.1] \times [ - 0.16,0.16] \,.
\end{equation}

The parameters specific to the spatial case are listed in Table~\ref{tab10}. These include the minimum allowable distances to the Earth and Moon, as well as the maximal admissible sizes for infeasible subdomains along each spatial direction. All other settings, such as the polynomial order and truncation threshold of the ADS, are identical to those used in the planar case (see Table~\ref{tab5}). Using the ADS process, a total of 7,662 subdomains are generated, among which 5,348 are identified as feasible based on the LDB method.

\begin{table}[!h]
    \caption{\label{tab10} User-defined parameters of the spatial case}
    \centering
    \begin{tabular}{lll}
    \hline\hline
        Parameter & Symbol & Value \\ \hline
        \multirow{2}{*}{\makecell[l]{Minimum allowable\\distance}} & $d_E^{\min }$ & $10^{-2}$ (LU) \\
         & $d_M^{\min }$ & $10^{-2}$ (LU) \\ \hline
        \multirow{4}{*}{\makecell[l]{Maximal size of\\infeasible subdomains}} & $\Delta x_0^{\max }$ & $10^{-2}$ (LU) \\
         & $\Delta \dot x_0^{\max }$ & $2\times10^{-2}$ (VU) \\
         & $\Delta z_0^{\max }$ & $1.25\times10^{-2}$ (LU) \\
         & $\Delta \dot z_0^{\max }$ & $4\times10^{-2}$ (VU) \\           
    \hline\hline
    \end{tabular}
\end{table}

Table~\ref{tab11} presents the fixed points obtained in the spatial case, all corresponding to 4-revolution POs. Since the constructed initial domain includes the $z=0$ plane, many planar solutions, such as the DRO, % (see Fig.~\ref{fig6a} and Fig.~\ref{fig15a}),
its multiple repetitions, and the P3DRO, % (see Fig.~\ref{fig6b}),
are also identified during the process. However, these planar solutions are manually excluded in the post-processing stage and are not listed in Table~\ref{tab11}.

\begin{table}[!h]
    \caption{\label{tab11} Fixed points obtained in the spatial case}
    \centering
    \begin{tabular}{lllll}
    \hline\hline
    \multicolumn{2}{l}{\multirow{2}{*}{Fixed points}} & \multicolumn{3}{l}{Revolution} \\ \cline{3-5}
    \multicolumn{2}{l}{} & 4 & 4 & 4 \\ \hline
    \multirow{6}{*}{\makecell[l]{Initial\\state}} & $x_0$ (LU) & 0.844996113719814 & 0.853531132520664 & 0.832978141490628 \\
     & $z_0$ (LU) & 0.0591969024456453 & 0.0588471252610341 & -0.00482324931511189 \\
     & $\dot{x}_0$ (VU) & 0.00843493943755992 & 0 & -0.0472506344683627 \\
     & $\dot{y}_0$ (VU) & 0.455607866485485 & 0.468318709618996 & 0.464988236479847 \\
     & $\dot{z}_0$ (VU) & 0.105436859672317 & 0 & -0.134360833873340 \\  \hline
    \multicolumn{2}{l}{Period (TU)} & 10.2121127631777 & 10.1659941262732 & 10.2220099597473 \\ \hline
    \multicolumn{2}{l}{Figure} & Fig.~\ref{fig17a} & Fig.~\ref{fig17b} & Fig.~\ref{fig17c}     \\ \hline\hline 
    \end{tabular}
\end{table}

\begin{figure}[!h]
    \centering
    \subfigure[First PO]{
        \label{fig17a}
        \includegraphics[width=0.75\linewidth]{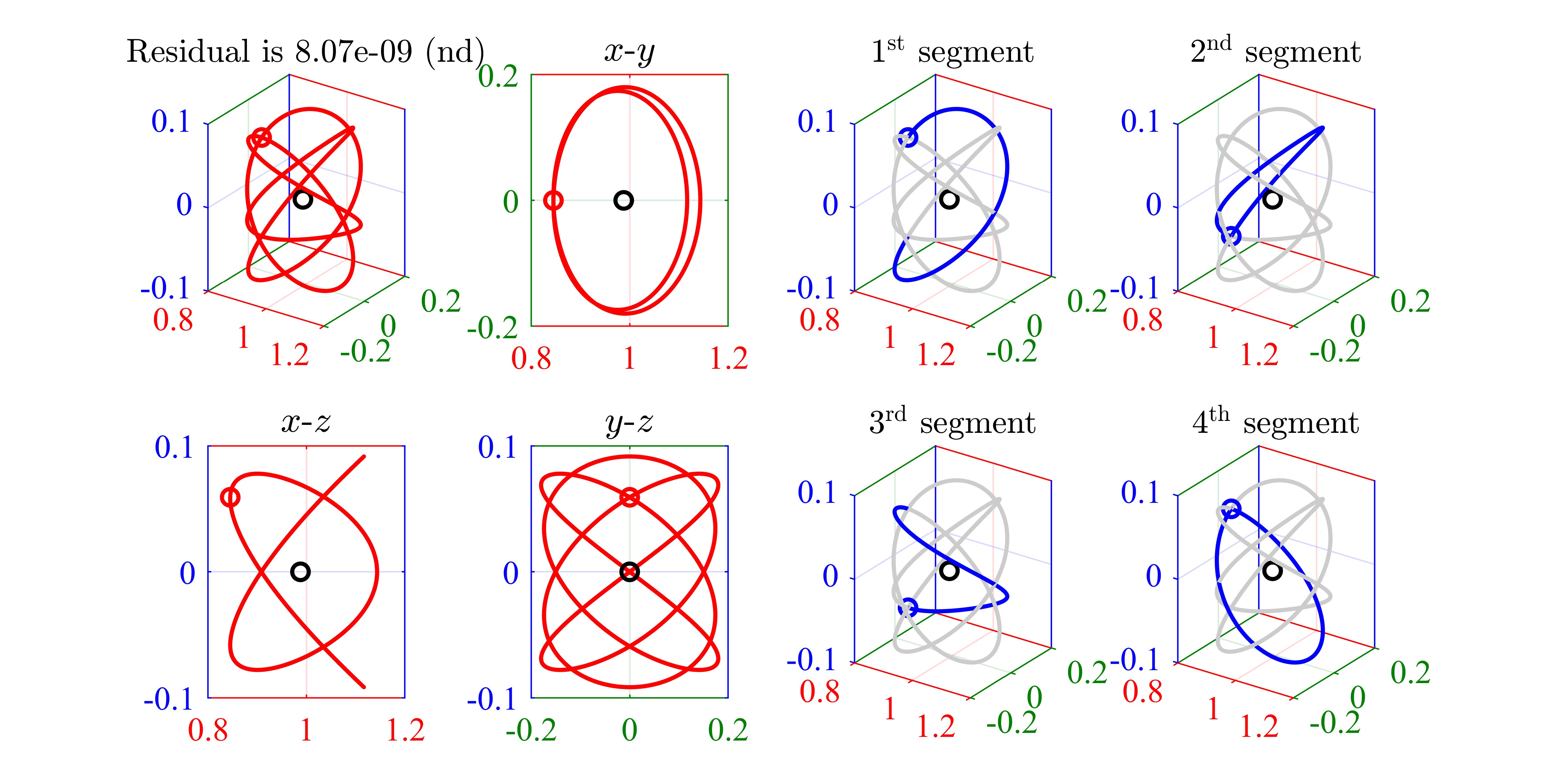}
    }
    \subfigure[Second PO]{
        \label{fig17b}
        \includegraphics[width=0.75\linewidth]{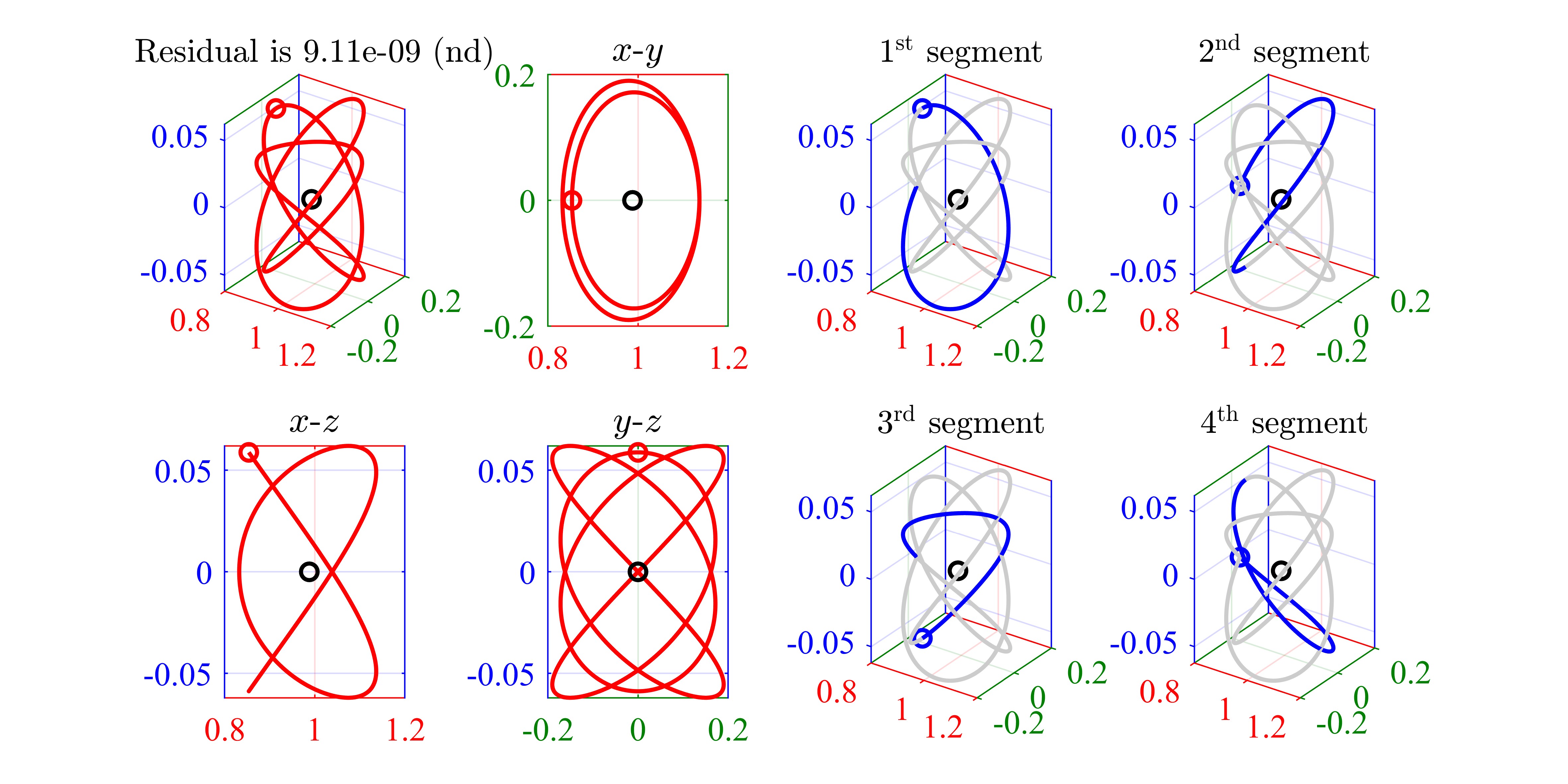}
    }
    \subfigure[Third PO]{
        \label{fig17c}
        \includegraphics[width=0.75\linewidth]{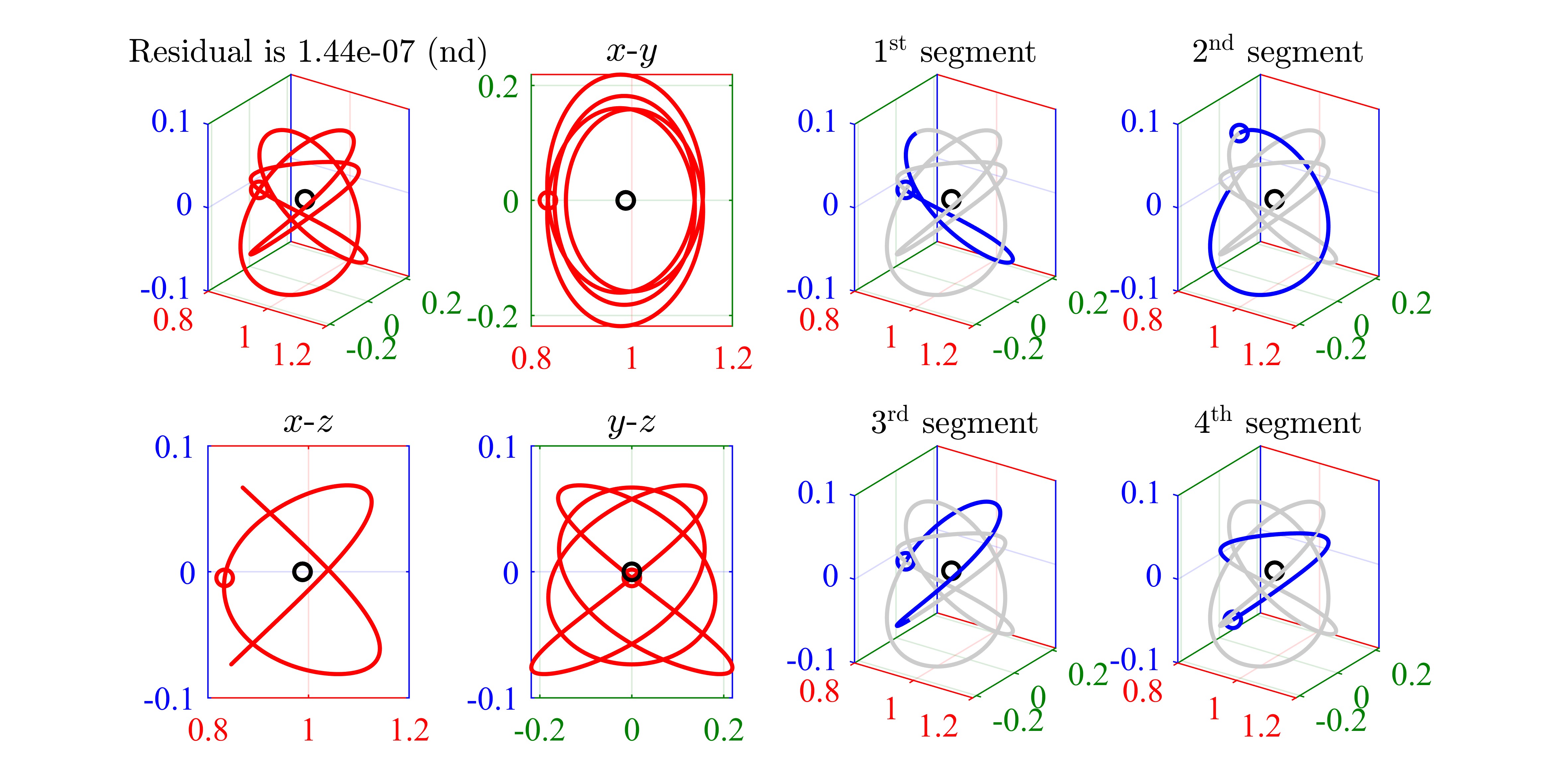}
    }
    \caption{\label{fig17} POs in the spatial case ($n=4$). The $x$-, $y$-, and $z$-axes are shown in red, green, and blue, respectively.}
\end{figure}

The POs corresponding to the fixed points in Table~\ref{tab11} are illustrated in Fig~\ref{fig17a}, Fig~\ref{fig17b}, and Fig~\ref{fig17c}. For clarity, the $x$-, $y$-, and $z$-axes in Fig~\ref{fig17} are rendered in red, green, and blue, respectively.
As shown in Fig~\ref{fig17a} and Fig~\ref{fig17b}, the first two 4-revolution POs exhibit symmetry with respect to the $z=0$ plane, while the third PO in Fig~\ref{fig17c} lacks this symmetry. A symmetric counterpart of this third spatial PO has also been identified; however, it is omitted from Table~\ref{tab11} for brevity. This counterpart can be easily recovered by reversing the signs of the ${z_0}$ and ${\dot z_0}$ components of the initial state listed in Table~\ref{tab11}. This type of symmetry is analogous to the northern and southern families of halo orbits. In addition, it is worth noting that the PO illustrated in Fig~\ref{fig17b} contains a point that intersects the Poincaré section at $z=0$. Consequently, if the spatial initial domain were constructed by enforcing $z>0$, this particular solution would not have been captured. This highlights the importance of carefully selecting the initial domain formulation, especially in spatial problems where certain periodic solutions may lie exactly on or cross the symmetry plane.

Additionally, we have also tested for 5-revolution and 6-revolution POs in the spatial case, but did not identify any valid fixed points. As a result, the search was not extended to higher revolution counts. 

%%%%%%%%%%%%%%%%%%%%%%%%%%%%%%%%%%%%%%%%%%%%%%%%%%%%%%%%%%%%%%%%%%%%%%%%%%%%%%%%%%%%%%%%%%%
%%%%%%%%%%%%%%%%%%%%%%%%%%%%%%%%%%%%%%%%%%%%%%%%%%%%%%%%%%%%%%%%%%%%%%%%%%%%%%%%%%%%%%%%%%%
\section{Conclusion} \label{sec:Conclusion}
A two-stage polynomial optimization framework has been proposed for identifying multiple-revolution fixed points in the circular restricted three-body problem (CRTBP), leveraging high-order transfer maps (HOTMs) constructed via differential algebra (DA) and enhanced by automatic domain splitting (ADS). The first stage efficiently selects combinable subdomain sequences from the ADS, whereas the second stage refines and locates fixed points along valid trajectories. Both stages formulate polynomial optimization problems based on the constructed HOTMs and solve them using a recursive polynomial optimization method to ensure accuracy and efficiency. This proposed framework has been successfully applied to both planar and spatial Earth-Moon CRTBP scenarios, demonstrating its capability to uncover fixed points up to nine revolutions in the planar case and four revolutions in the spatial case. Novel orbit families are discovered in the planar case and named P\emph{n}g’. They exhibit hybrid characteristics of Lyapunov and DRO-type motion, which may offer practical applications in trajectory design. The fixed-point computation method requires only a single construction of the HOTM, enabling scalable fixed-point search across arbitrary revolution counts. Simulation results confirm the effectiveness of the approach, with manageable polynomial approximation errors and a theoretical capability to discover all fixed points within the given domain.

\section*{Acknowledgments}
The first part of this work, concerning the development of high-order transfer map (HOTM) techniques and their application to the study of chaotic cislunar dynamics within the Circular Restricted Three-Body Problem (CRTBP), was carried out in collaboration with Prof. Hinke M. Osinga and Prof. Bernd Krauskopf from the Department of Mathematics at the University of Auckland, New Zealand. Their expertise and valuable contributions were instrumental to the ideation of this technique and its application in the broader context of dynamical systems.

This work was supported by the National Natural Science Foundation of China (No. 12150008, No. 124B2049, No. 62394353) and the Changjiang Scholars Program (No. T2023191). Xingyu Zhou is grateful for the financial support provided by the China Scholarship Council (No. 202406030186) and the China Association for Science and Technology (CAST) Young Talent Support Program (No. 156-O-590-0000198-7).

\bibliography{sample}

\end{document}